\documentclass[fleqn,usenatbib]{mnras}

\usepackage{newtxtext,newtxmath}

\usepackage[T1]{fontenc}

\DeclareRobustCommand{\VAN}[3]{#2}
\let\VANthebibliography\thebibliography
\def\thebibliography{\DeclareRobustCommand{\VAN}[3]{##3}\VANthebibliography}

\usepackage{graphicx}	
\usepackage{amsmath}	
\usepackage{booktabs}
\usepackage{multirow}
\usepackage{siunitx}
\usepackage{textgreek}
\usepackage[export]{adjustbox}
\usepackage{placeins}

\usepackage{enumitem} 
\setlist[enumerate]{labelindent=0pt, leftmargin=*, widest=iii, align=right} 
\setlist[itemize]{labelindent=0.5em,leftmargin=*, align=right}

\renewcommand{\epsilon}{\varepsilon}
\newcommand\smaller[2][0.85]{{\scalefont{#1}#2}}
\newcommand\dppp{\textsc{dp}\smaller[0.76]{3}}

\title[First upper limits on the 21-cm power spectrum at $z=9.16$ from LOFAR 3C\texorpdfstring{\,}{}196]{First upper limits on the 21-cm signal power spectrum of neutral hydrogen at $z=9.16$ from the LOFAR 3C\texorpdfstring{\,}{}196 field}

\author[E. Ceccotti et al.]{E.\ Ceccotti$^{1,2}$\thanks{E-mail: emilio.ceccotti@inaf.it},
A.~R.\ Offringa$^{3,1}$,
F.~G.\ Mertens$^{4}$,
L.~V.~E.\ Koopmans$^{1}$,
S.~Munshi$^{1}$,
J.~K.~Chege$^{1,3}$,
\newauthor
A.~Acharya$^{5}$,
S.~A.~Brackenhoff$^{1}$,
E.~Chapman$^{6}$,
B.~Ciardi$^{5}$,
R.~Ghara$^{7}$,
S.~Ghosh$^{1}$,
S.~K.~Giri$^{8,9}$,
\newauthor
C.~H\"{o}fer$^{1}$,
I.~Hothi$^{10}$,
G.~Mellema$^{11}$,
M.~Mevius$^{3,1}$,
V.~N.~Pandey$^{3}$
and S.~Zaroubi$^{12,1,5}$
\\
$^{1}$ Kapteyn Astronomical Institute, University of Groningen, PO Box 800, 9700 AV Groningen, The Netherlands\\ 
$^{2}$ INAF -- Istituto di Radioastronomia, Via P.~Gobetti 101, 40129 Bologna, Italy\\
$^{3}$ ASTRON, PO Box 2, 7990 AA Dwingeloo, The Netherlands\\
$^{4}$ LERMA, Observatoire de Paris, PSL Research University, CNRS, Sorbonne Université, F-75014 Paris, France\\
$^{5}$ Max-Planck Institute for Astrophysics, Karl-Schwarzschild-Straße 1, 85748 Garching, Germany\\
$^{6}$ School of Physics and Astronomy, The University of Nottingham, University Park, Nottingham, NG7 2RD, UK\\
$^{7}$ Department of Physical Sciences, Indian Institute of Science Education and Research Kolkata, Mohanpur, WB 741 246, India\\
$^{8}$ Van Swinderen Institute for Particle Physics and Gravity, University of Groningen, Nĳenborgh 4, 9747 AG Groningen, The Netherlands\\
$^{9}$ Nordita, KTH Royal Institute of Technology and Stockholm University, Hannes Alfvéns väg 12, SE-106 91 Stockholm, Sweden\\
$^{10}$ Laboratoire de Physique de l'ENS, ENS, Universit\'{e} PSL, CNRS, Sorbonne Universit\'{e}, Universit\'{e}e Paris Cit\'{e}, 75005 Paris, France\\
$^{11}$ The Oskar Klein Centre, Department of Astronomy, Stockholm University, AlbaNova, SE-10691 Stockholm, Sweden\\
$^{12}$ Department of Natural Sciences, The Open University of Israel, 1 University Road, PO Box 808, Ra’anana 4353701, Israel
}

\date{Accepted XXX. Received YYY; in original form ZZZ}

\pubyear{\the\year{}}

\begin{document}
\label{firstpage}
\pagerange{\pageref{firstpage}--\pageref{lastpage}}
\maketitle

\begin{abstract}
The redshifted 21-cm signal of neutral hydrogen from the Epoch of Reionization (EoR) can potentially be detected using low-frequency radio instruments such as the Low-Frequency Array (LOFAR). So far, LOFAR upper limits on the 21-cm signal power spectrum have been published using a single target field: the North Celestial Pole (NCP). In this work, we analyse and provide upper limits for the 3C\,196 field, observed by LOFAR, with a strong ${\approx}80$\,Jy source in the centre. This field offers advantages such as higher sensitivity due to zenith-crossing observations and reduced geostationary radio-frequency interference, but also poses challenges due to the presence of the bright central source. After constructing a wide-field sky model, we process a single 6-hour night of 3C\,196 observations using direction-independent and direction-dependent calibration, followed by a residual foreground subtraction with a machine learned Gaussian process regression (ML-GPR). A bias correction is necessary to account for signal suppression in the GPR step. Still, even after this correction, the upper limits are a factor of two lower than previous single-night NCP results, with a lowest $2\sigma$ upper limit of $(146.61\,\text{mK})^2$ at $z = 9.16$ and $k=0.078\,h\,\text{cMpc}^{-1}$ (with $\text{d}k/k\approx 0.3$). The results also reveal an excess power, different in behaviour from that observed in the NCP field, suggesting a potential residual foreground origin. In future work, the use of multiple nights of 3C\,196 observations combined with improvements to sky modelling and ML-GPR to avoid the need for bias correction should provide tighter constraints per unit observing time than the NCP.
\end{abstract}

\begin{keywords}
cosmology: dark ages, reionization, first stars -- cosmology: observations -- techniques: interferometric -- methods: data analysis
\end{keywords}



\section{Introduction}

After cosmological recombination ($z\approx 1100$), hydrogen in the Universe became neutral and the Dark Ages began. During this era, matter density fluctuations grew under the influence of gravity, leading to the formation of the first stars and galaxies, marking the onset of the Cosmic Dawn (CD; $30\gtrsim z \gtrsim 15$). Ultraviolet and X-ray radiation from these first objects initially heated and then ionized the surrounding intergalactic medium (IGM). This final major transition of the Universe, known as the Epoch of Reionization (EoR; $15\gtrsim z \gtrsim 6$), marked the transformation of the IGM from largely neutral to largely ionized \citep[e.g.][]{barkana_loeb:2001,loeb_furlanetto:2013}.

Our understanding of the EoR has advanced by indirect probes such as spectra of high-redshift quasars \citep[e.g.][]{becker_etal:2001,fan_etal:2006,venemans_etal:2013,eilers_etal:2018,keller_etal:2024}, Cosmic Microwave Background optical depth \citep[e.g.][]{planck_coll:VI:2020}, and Lyman-$\alpha$ emitters \citep[e.g.][]{ouchi_etal:2010,konno_etal:2014,zheng_etal:2017,taylor_etal:2021,witstok_etal:2024}. Recent breakthroughs by the  James Webb Space Telescope have revealed an unexpectedly rich population of luminous objects at $z>10$ \citep[e.g.][]{bradley_etal:2023,finkelstein_etal:2023,carniani_etal:2024,mcleod_etal:2024,chemerynska_etal:2024}, challenging existing galaxy formation models \citep[e.g.][]{arrabal-haro_etal:2023,boylan-kolchin:2023,ferrara_etal:2023,mason_etal:2023}. However, uncertainties remain on the dominant ionizing sources, and these probes are unable to directly trace the reionization timeline. The most direct probe to map the evolution of the IGM during these early epochs is the redshifted 21-cm line from the hyperfine spin-flip transition of neutral hydrogen \citep[see e.g.][for some reviews]{furlanetto_etal:2006,pritchard_loeb:2012,liu_shaw:2020}.

The 21-cm line emitted at $z>6$ is observed at wavelengths longer than 1.5\,m, and hence we need low-frequency radio instruments to detect it. Current instruments lack the sensitivity to directly image the 21-cm signal, aiming instead for a statistical detection of its spatial fluctuations by measuring its power spectrum. Instruments such as LOFAR\footnote{Low-Frequency Array, \url{https://www.astron.nl/telescopes/lofar}} \citep{vanhaarlem_etal:2013,mertens_etal:2020}, MWA\footnote{Murchison Widefield Array, \url{http://www.mwatelescope.org}} \citep{lonsdale_etal:2009,trott_etal:2020,kolopanis_etal:2023}, NenuFAR\footnote{New Extension in Nançay Upgrading LOFAR, \url{https://nenufar.obs-nancay.fr}} \citep{zarka_etal:2012,munshi_etal:2024:ul}, GMRT\footnote{Giant Metrewave Radio Telescope, \url{https://www.gmrt.ncra.tifr.res.in}} \citep{paciga_etal:2013,gupta_etal:2017}, and HERA\footnote{Hydrogen Epoch of Reionization Array, \url{https://reionization.org/}} \citep{deboer_etal:2017,hera_collaboration:2023} have set increasingly stringent upper limits on the 21-cm power spectrum at different redshifts and scales, both at the CD and EoR. The upcoming SKA-Low\footnote{Square Kilometre Array, \url{https://www.skao.int/en/explore/telescopes}} \citep{dewdney_etal:2009,koopmans_etal:2015} is expected to have the sensitivity to make a direct detection and image the 21-cm signal \citep[e.g.][]{giri_etal:2018a,giri_etal:2018b,bianco_etal:2024}. 

All the aforementioned instruments face many challenges in detecting the 21-cm signal, which is a few orders of magnitude fainter than the astrophysical foregrounds, namely extra-galactic and Galactic emission. Because these foregrounds have smooth frequency spectra, being dominated by synchrotron radiation at low-frequencies, they can be separated from the 21-cm signal, which fluctuates rapidly, being a hyperfine spectral line \citep{jelic_etal:2008,bowman_etal:2009,bernardi_etal:2009,bernardi_etal:2010}. Foreground mitigation can be done with different techniques, such as foreground avoidance, where the Fourier modes dominated by foreground emission are directly discarded \citep[e.g.][]{parsons_backer:2009,parsons_etal:2012,thyagarajan_etal:2015}, and foreground subtraction, where sky emission is subtracted from the data by accurate modelling \citep[e.g.][]{patil_etal:2017,mertens_etal:2020}. However, radio interferometers are intrinsically chromatic and introduce spectral structures to the foregrounds, resulting in `mode-mixing' that makes the signal separation harder \citep{morales_etal:2012,morales_etal:2019}. Errors can also be caused by inaccurate calibration \citep{barry_etal:2016,beardsley_etal:2016,patil_etal:2016,ewall-wice_etal:2017,mazumder_etal:2022,gorce_etal:2023,ceccotti_etal:2025}, polarization leakage \citep{jelic_etal:2010, spinelli_etal:2018,cunnington_etal:2021}, ionospheric effects \citep{koopmans:2010,vedantham_koopmans:2016,mevius_etal:2016,brackenhoff_etal:2024}, inaccurate primary beam model \citep{gehlot_etal:2021,chokshi_etal:2024}, gridding of visibilities \citep{offringa_etal:2019b}, as well as mutual coupling \citep{kern_etal:2020, kolopanis_etal:2023,rath_etal:2024} and low-level radio-frequency interference \citep[RFI;][]{offringa_etal:2019a,wilensky_etal:2019}. 

By addressing many of these challenges, the LOFAR-EoR Key Science Project (KSP) has made continuous progress in recent years. The KSP focuses on two main sky fields: the North Celestial Pole (NCP) field and the 3C\,196 field, both of which are relatively cold spots in the Milky Way \citep{bernardi_etal:2010}. The two fields have contrasting characteristics, some of which are described by \citet{yatawatta:2013}. The main difference is the presence of a very bright FR-II source at the centre of the 3C\,196 field, which can help the calibration by providing a high signal-to-noise ratio but may leave residuals after subtraction. Such a bright source is absent in the NCP field. Another difference is the elevation: with a declination of approximately $48^\circ$, 3C\,196 passes close to the zenith, whereas the NCP has a fixed elevation of $53^\circ$. Consequently, the thermal noise in the 3C\,196 direction is expected to be lower than in the NCP direction because the gain of the LOFAR dipole beam is maximum at zenith. On the other hand, the NCP field can be observed throughout the year and with a circular $u\varv$-coverage, as its pointing direction is fixed and parallel to the Earth rotation axis. This also introduces another difference, related to geostationary RFI (i.e.\ RFI sources that do not move with respect to the array), whose signals accumulate coherently in the NCP field, making it more challenging to remove \citep{munshi_etal:2025:rfi}. Being closer to the Galactic plane than the NCP, the 3C\,196 field exhibits bright linearly polarized structures, which can again contaminate the 21-cm power spectrum if not properly accounted for \citep{jelic_etal:2015}. This presents a challenge, as polarized emission is primarily attributed to diffuse Galactic emission and is affected by Faraday rotation, which are currently not accounted for in the LOFAR-EoR 21-cm signal processing pipelines. Nevertheless, \citet{asad_etal:2015} demonstrated that such contamination remains at the level of the expected EoR signal, or below it if mildly subtracted, making it negligible for the current upper limits.

Both fields have been observed for thousands of hours, providing the sensitivity needed to detect the 21-cm signal in the absence of systematics. However, most of the calibration and analysis efforts have focused on the NCP field, where an extensive sky model was made by \citet{yatawatta:2013} at 150\,MHz. This model was used by \citet{patil_etal:2017} to get the first upper limits in the redshift range $z=7.9{-}10.6$ from 13\,h of data. The wide-field sky model was used for the initial calibration and then split into multiple directions to solve for direction-dependent (DD) station gains using the \textsc{sagecal}\footnote{\url{https://sagecal.sourceforge.net/}} code \citep{yatawatta:2016}. The foreground subtraction was then performed by applying the DD solutions to the sky model, and the residual foreground emission was dealt with Generalized Morphological Component Analysis \citep[GMCA;][]{bobin_etal:2007,bobin_etal:2008,chapman_etal:2013}. Using similar but improved processing pipeline on 141\,h of data (12 nights) and replacing GMCA with the Gaussian process regression \citep[GPR;][]{mertens_etal:2018}, \citet{mertens_etal:2020} have been able to set deeper upper limits at $z\approx 9.1$ on the spherical power spectrum, namely $\Delta_{21}^2 < (72.86\,\text{mK})^2$ at $k=0.075\,h\,\text{cMpc}^{-1}$ (with $\text{d}k/k\approx 0.3$). 

These results already set some constraints on the physics of the EoR \citep[e.g.][]{ghara_etal:2020,mondal_etal:2020,greig_etal:2021}, but the upper limits were systematics limited rather than thermal noise limited. Consequently, increasing the observing time does not lower the upper limits if the excess power is not incoherent in time like the thermal noise. Understanding the origin of this excess is therefore crucial for mitigating or avoiding it. Recent studies on the NCP field have shown that ionospheric effects are unlikely to be the primary contributors \citep{gan_etal:2022,brackenhoff_etal:2024}, although bright, distant sources such as Cassiopeia\,A (Cas\,A) and Cygnus\,A (Cyg\,A) might play a role, especially in combination with incorrect primary beam models and DD gain errors \citep{gan_etal:2022,ceccotti_etal:2025,brackenhoff_etal:2025}. Despite these effects, GPR \citep{mertens_etal:2018} has demonstrated competitive performance even in the presence of such an excess \citep{hothi_etal:2021}, although this methodology might suffer from biases and signal suppression, as highlighted by \citet{kern_liu:2021}. To address these effects, a machine learning-enhanced GPR framework was developed by \citet{mertens_etal:2024}, incorporating parametrized covariance functions derived from simulations. This framework has been successfully tested on both simulations \citep{acharya_etal:2023} and real LOFAR data \citep{acharya_etal:2024}. The better understanding of the excess power has led to new upper limits at $z\approx8.3$, $9.1$, and $10.1$ from the NCP field \citep{mertens_etal:2025}, deeper by a factor of $2{-}4$, depending on the $k$-mode, at $z\approx9.1$ compared to \citet{mertens_etal:2020}, using nearly the dame data sets.

While most efforts have focused on the NCP field, investigating the second deep field, centred on 3C\,196, could provide new insights into the origin of the excess power. Low-band observations of 3C\,196 in the $56{-}70\,\text{MHz}$ range (i.e.\ $19 \lesssim z \lesssim 24$) were analysed by \citet{gehlot_etal:2018}, revealing a significant excess power across all baseline lengths. Similar to the NCP case, potential sources for this excess have been proposed, such as an incomplete sky model, ionospheric effects, and calibration errors, but no extensive investigation has been carried out in the EoR frequency window. Additionally, it has been demonstrated that comparing results from different fields is crucial for better understanding the sources of systematics \citep[e.g.][]{trott_etal:2020,rahimi_etal:2021,abdurashidova_etal:2022}.

In this paper, we present the first 21-cm signal power spectrum upper limits at $z \approx 9.1$ from LOFAR observations of the 3C\,196 field. As discussed earlier, the 3C\,196 field poses different challenges compared to the NCP, particularly due to the presence of a bright central source, which makes its processing and analysis more complex. Building on the experience gained by the LOFAR-EoR team from NCP observations, we developed a processing pipeline, tailored for this field, incorporating the latest software and techniques. With this pipeline, we analysed a single 6-hour night of observed data resulting in a 21-cm signal power spectrum independent from the NCP field.

The paper is organized as follows. In Section~\ref{sec:3c196-obs}, we introduce the 3C\,196 dataset, which is used to extract a wide-field sky model and estimate the 21-cm upper limits. In Section~\ref{sec:sky-modelling}, we describe the process of obtaining the sky model for the field, with several processing steps mirroring those used in the new EoR pipeline, detailed in Section~\ref{sec:3c196-processing}. In Section~\ref{sec:gpr}, we discuss the GPR method and present the resulting power spectra. In Section~\ref{sec:validation}, we focus on the validation of the 21-cm signal processing pipeline, particularly the GPR technique. The results and final upper limits derived from the 3C\,196 field are presented in Section~\ref{sec:results}. Finally, in Section~\ref{sec:conclusions}, we summarise the findings and draw the conclusions. Throughout this paper, we used a flat $\Lambda$CDM cosmology consistent with the \citet{planck_coll:XIII:2016} results, similar to \citet{mertens_etal:2020}.

\section{LOFAR-HBA observations of the 3C\texorpdfstring{\,}{}196 field}\label{sec:3c196-obs}

LOFAR \citep{vanhaarlem_etal:2013} is a low-frequency radio interferometer constituted by stations spread across the Netherlands and Europe. It currently consists of 24 core stations (CS; max baseline $\approx4$\,km) and 14 remote stations (RS; max baseline $\approx 120$\,km) in the Netherlands, forming the Dutch array, and 14 international stations (max baseline $\approx2000$\,km) spread across Europe. Every station is organized into two types of antenna sets observing different frequency ranges: Low-Band Antennas (LBA) for $10{-}90\,\text{MHz}$ and High-Band Antennas (HBA) for $110{-}250\,\text{MHz}$. Each CS has two HBA sub-stations (HBA0 and HBA1) that can operate independently (`HBA Dual' mode) for a better $u\varv$-coverage. Each station acts as a phased array that can track a phase centre on the sky via beam-forming inside the dipole beam.

Since its first observations in 2011, the LOFAR-EoR KSP has focused on observing two main windows on the sky using the HBA system: the North Celestial Pole \citep{patil_etal:2017,mertens_etal:2020,mertens_etal:2025} and the 3C\,196 fields \citet{gehlot_etal:2018}. Compared to the NCP \citep[see][]{yatawatta:2013}, the 3C\,196 field presents advantages and disadvantages. Firstly, with a flux density of about 83\,Jy at 150\,MHz \citep{scaife_heald:2012}, 3C\,196 requires a very accurate model to minimize its residuals during the foreground subtraction step (see Section~\ref{sec:proc:dd-cal}). Once such a model is in place, having a bright source in the centre makes the direction-independent calibration easier because 3C\,196 dominates the observed visibilities. Another challenge is that, while the NCP phase centre is fixed in the sky and is observable by LOFAR at the same elevation throughout the year, 3C\,196 moves due to the Earth rotation and is observable only for half of the year. A benefit is that it can be observed close to the zenith, having a declination of $48.2^\circ$ (whereas LOFAR core is at a latitude of ${\approx}53^\circ$), where the dipole beam suppression is more limited compared to the NCP. This results in a higher sensitivity (approximately a factor of 1.5 more than for the NCP field). The fixed pointing direction of the NCP also leads to geostationary RFI to add up coherently, having the same fringe speed as sources in the target field, while this is not the case for a tracking field such as the one centred on 3C\,196 \citep{munshi_etal:2025:rfi}. 

\subsection{Selected data set}\label{sec:dataset}

In this work, we analysed a single night of 6\,h LOFAR-HBA observations taken in 2014 (with elevation ranging from $61^\circ$ to $85^\circ$, peaking at the midpoint of the observation). The recorded data consists of 380 sub-bands (SBs) spanning 115--189\,MHz, where each SB is 195.3\,kHz wide, with a frequency resolution of 3.05\,kHz (i.e.\ 64 channels per SB) and an integration time of 2\,s. However, because of the poly-phase filter \citep{brentjens_mol:2018}, the four edge channels of each SB are affected by aliasing and were flagged before starting any processing. This reduces the SB width to 183.1\,kHz. The data were flagged with \textsc{aoflagger} \citep{offringa_etal:2012}, and averaged to 5 channels (36.6\,kHz each) per SB and 4\,s for archival purposes. The data were stored uncompressed because \textsc{dysco} compression \citep{offringa:2016,chege_etal:2024} was not yet implemented at that time. The resulting data set is the starting point of this work and will be referred to as the initial data set. Further observational details are reported in Table~\ref{tab:obs-details}. All the processing described in this paper has been performed on the `Dawn' high-performance GPU computing cluster \citep{pandey_etal:2020} at the University of Groningen.

\begin{table}
\caption{Observational details of the initial 3C\,196 data set used in this work.}
\label{tab:obs-details}
\begin{tabular}{lc}
\toprule
Parameter & Value \\
\midrule
Telescope & LOFAR HBA \\
Project code & LC3\_028 \\
Observation ID & L253456 \\
Antenna configuration & HBA Dual Inner \\
Number of stations & $61$ ($\text{CS} + \text{RS})^a$  \\
Phase centre (J2000): & {} \\
{ }{ } Right Ascension & $08^\text{h}13^\text{m}36\rlap{.}^\text{s}07$ \\
{ }{ } Declination & $+48^\circ13'02\rlap{.}''58$ \\
Obs.\ start time (UTC) & 2014 Dec 02; 23:58:39 \\
Frequency range: & {} \\
{ }{ } Full data set & $115.0{-}189.1\,\text{MHz}$ \\
{ }{ } Redshift bin & $134.2{-}147.1\,\text{MHz}$ \\
Duration of observation & 6\,h \\
Time resolution & 4.0\,s \\
Frequency resolution & 36.6\,kHz\\
\bottomrule
\end{tabular}
\vspace{1ex}

\hspace{1ex}{\footnotesize $^a$ RS310 did not participate in this observation. \par}
\end{table}

\section{Sky modelling}\label{sec:sky-modelling}

To calibrate our data, a sky model of the 3C\,196 field is required. While for a direction-independent (DI) calibration just using the bright 3C\,196 source might be enough, for the direction-dependent calibration and subtraction required before the power spectrum estimation, we need an extensive, wide-field model of the surrounding field. In the following, we describe the processing steps used to produce such a sky model.

We used the same data set presented in Section~\ref{sec:dataset}, but we selected 246 SBs spanning the 120--168\,MHz frequency range. The choice of not using the full 75\,MHz bandwidth comes from two main reasons: (i) the band edges are mostly affected by RFI, and (ii) to speed up calibration and imaging. Because the data set has already been flagged before archiving, we can start the processing directly with the DI calibration, described in the next section. 


\subsection{Baseline selection}\label{sec:mod:baselines}

\begin{figure}
    \centering
    \includegraphics[width=1\columnwidth]{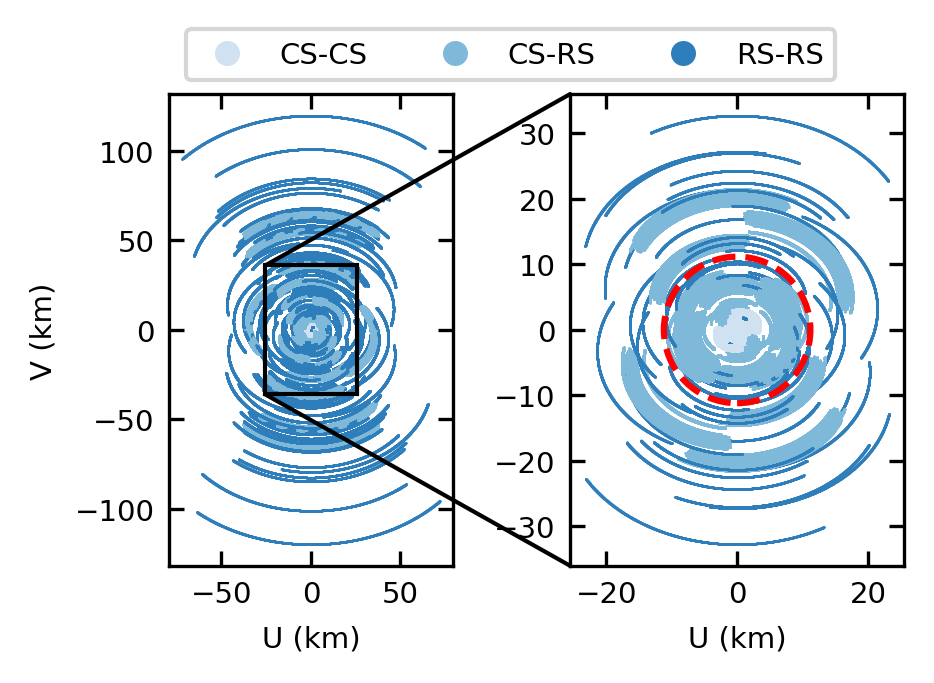}
    \caption{Baseline coverage in the $u\varv$-plane for the LOFAR-HBA observation of the 3C\,196 field, before (left, zoomed out) and after (right, zoomed in) the flagging performed in the pre-processing step (see  Section~\ref{sec:mod:baselines} and Section~\ref{sec:proc:pre-proc}). The antenna pairs are shown in different colours: core-to-core station baselines in light blue, core-to-remote (and remote-to-core) station baselines in medium blue, and remote-to-remote baselines in dark blue. In the right panel, the red circle marks the $5000\lambda$ limit that is usually applied during the DI calibration of the NCP field.}
    \label{fig:3C196_uvcov}
\end{figure}

In general, when we are interested in a single, dominant source at the phase centre, DI calibration can be performed with all the available baselines, resulting in the maximum signal-to-noise ratio for the station gains. In this case, most of the DD effects can be neglected. In our case, we want to extract a model of a few degrees in spatial extent from the target field, which means that we have to minimize the DD effects, such as ionospheric distortion and smearing of distant sources \citep[e.g.][]{patil_etal:2016, vedantham_koopmans:2015}. This is done by removing the longest baselines, applying an outer $u\varv$-cut. By removing only the longest baselines, most stations still have sufficient baselines to obtain accurate gain solutions. The challenge here is that we still want long baselines to get a sky model with a high enough spatial resolution. We found that using baselines shorter than ${\approx}30\,\text{km}$ (i.e.\ ${\approx}15\,000\lambda$ at 150\,MHz) is a good compromise between minimizing ionospheric distortion, given a diffractive scale of about 15\,km for our observed night, and maximising spatial resolution, whose full width at half maximum (FWHM) is approximately 14\,arcsec. Instead of cutting in the $u\varv$-plane, we removed all the baselines that had a physical length larger than 30\,km. This avoids having gaps in the  $u\varv$-tracks for some baselines during the time synthesis, due to the ellipticity of $u\varv$-tracks. We also removed a few RS that resulted in just one baseline, which is not enough to obtain reliable gain solutions. Fig.~\ref{fig:3C196_uvcov} shows the $u\varv$-coverage before (left panel, zoomed out) and after such baseline selection (right panel, zoomed in). Furthermore, we also removed the intra-station baselines (HBA0-HBA1) of the core array, because of possible correlated RFI generated inside the shared electronics cabinet. This baseline selection is applied to the data set before DI calibration for the sky modelling part and is part of the pre-processing step of the EoR pipeline, later described in Section~\ref{sec:proc:pre-proc}. Additionally, a lower $u\varv$-cut of $50\lambda$ is applied throughout the entire processing to avoid poor $u\varv$-coverage and strong unmodelled diffuse emission \citep{patil_etal:2017}.

\subsection{Direction-independent calibration}\label{sec:mod:di-cal}

\begin{figure}
    \centering
    \includegraphics[width=1\columnwidth]{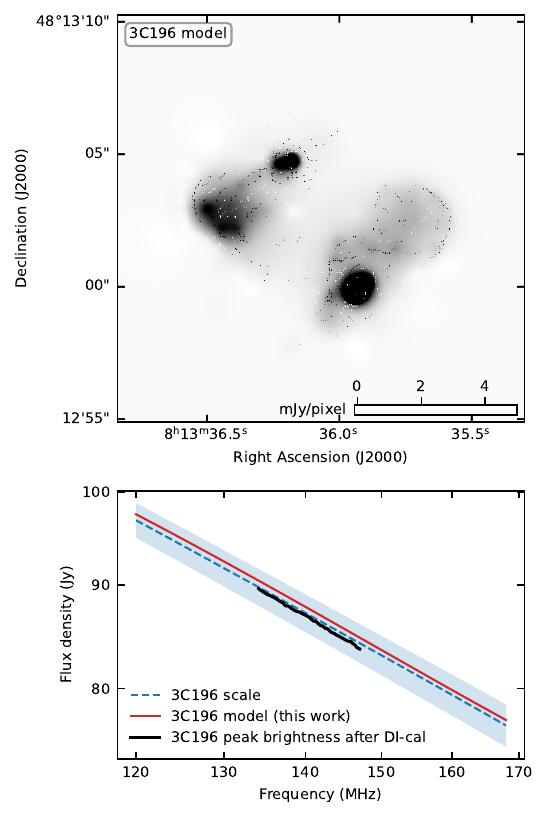}
    \caption{Rendered model of the 3C\,196 high-resolution model at 150\,MHz (top) and the flux density over the 120--168\,MHz range (bottom). The flux estimated by \citet{scaife_heald:2012} is plotted with the dashed blue line, with the $1\sigma$ uncertainties as the blue shaded area. The flux of the 3C\,196 model used in this work is shown with the red line. We also show the peak brightness of the source at each frequency of the $z$-bin, extracted from the dirty images after the DI-calibration of the EoR pipeline (Section~\ref{sec:proc:di-cal}), with the black line.}
    \label{fig:3C196_model_flux_afterDI}
\end{figure}

The DI calibration is performed by using a high-resolution, multi-scale LOFAR-HBA model of 3C\,196, shown in the top panel of Fig.~\ref{fig:3C196_model_flux_afterDI}. 3C\,196 is a compact source approximately 10\,arcsec in size, so LOFAR international stations were used to make such a model, which is described by 1812 components (634 point sources and 1178 Gaussians). The total flux agrees with the \citet{scaife_heald:2012} scale, as shown in the bottom panel of Fig.~\ref{fig:3C196_model_flux_afterDI}. To process and calibrate the data, we used \dppp\footnote{\url{https://dp3.readthedocs.io/}} \citep[Default Pre-Processing Pipeline;][]{vandiepen_etal:2018}, which collects many tools to perform different operations. DI calibration was divided in two steps, similar to the 21-cm signal processing pipeline that we will describe in Section~\ref{sec:3c196-processing}: (i) a spectrally smooth calibration, and (ii) a bandpass calibration. 

The main goal of the spectrally smooth calibration is to correct for high-temporal effects in the station gains. Using a spectral constraint helps keep the solutions stable over frequency while solving at the highest time resolution possible. This initial calibration was performed with the \textsc{ddecal} tool \citep[for an extensive description, see][]{brackenhoff_etal:2025}, which applies a Gaussian smoothing kernel at every iteration to enforce spectral smoothness in the solutions. In our case, a kernel of 1\,MHz width was used. We solved both amplitudes and phases of the full Jones matrix (a $2\times2$ complex matrix describing the effect of a station with orthogonal polarizations) for each SB (183.1\,kHz) and each time interval (4\,s). It is important to solve for a full Jones matrix (i.e.\ four complex gains) to correct any polarization leakage. Sometimes leakage from one Stokes~mode to another can occur because of the primary beam, as we will show in Section~\ref{sec:proc:dd-cal}. During calibration, the primary beam model response is applied to the 3C\,196 model to obtain apparent visibilities. By using the beam model throughout the processing, we can later extract intrinsic flux values of the entire field. In \dppp, the primary beam model includes both the element beam, which describes the directional sensitivity by combining the dipole beams in a single tile, and the array factor, which characterises the actual field of view of a station \citep{vanhaarlem_etal:2013}. These two beams are independently calculated and can be multiplied to obtain the full LOFAR primary beam. The array factor is negligible in our case, as the 3C\,196 is compact and in the centre of the field, making the array factor one at its position. However, since the field is not observed at zenith, the element beam is time-dependent and not unity. The response is approximately 90 per cent at the meridian. After calibration, the gain solutions were applied to the data.

After correcting for fast effects, the fine frequency response of the stations must be calibrated. For this bandpass calibration, we used the \textsc{gaincal} algorithm of \dppp\ for finding solutions \citep{mitchell_etal:2008,salvini_wijnholds:2014}. We solved for each channel (36.6\,kHz) over a time interval of 1\,h to average out the temporal behaviour of the instrument and obtain reliable spectral corrections. To reduce the number of degrees of freedom, we solved only for the diagonal elements of the Jones matrix. The bandpass solutions were used to subtract 3C\,196 from the data set, before applying them to the residual visibilities. Removing the central source is essential for the next imaging step, where the wide-field sky model will be extracted. Deconvolution struggles to handle the sidelobes of such a bright source, but our 3C\,196 model, which was built using very long baselines, is more spatially accurate than what imaging can reproduce from our data set, which includes Dutch baselines only. This allowed us to better remove the 3C\,196 sidelobes and achieve a more complete and accurate sky model for the rest of the field. The calibrated data were finally corrected for the element beam in the visibility space by scaling the data to have the correct flux density at the phase centre. To extract an intrinsic model of the 3C\,196 field, we applied the more rapidly spatially varying beam (i.e.\ the array factor term) during the imaging step, which is described in the next section.

\subsection{Imaging}

\begin{figure*}
    \centering
    \includegraphics[width=0.95\textwidth]{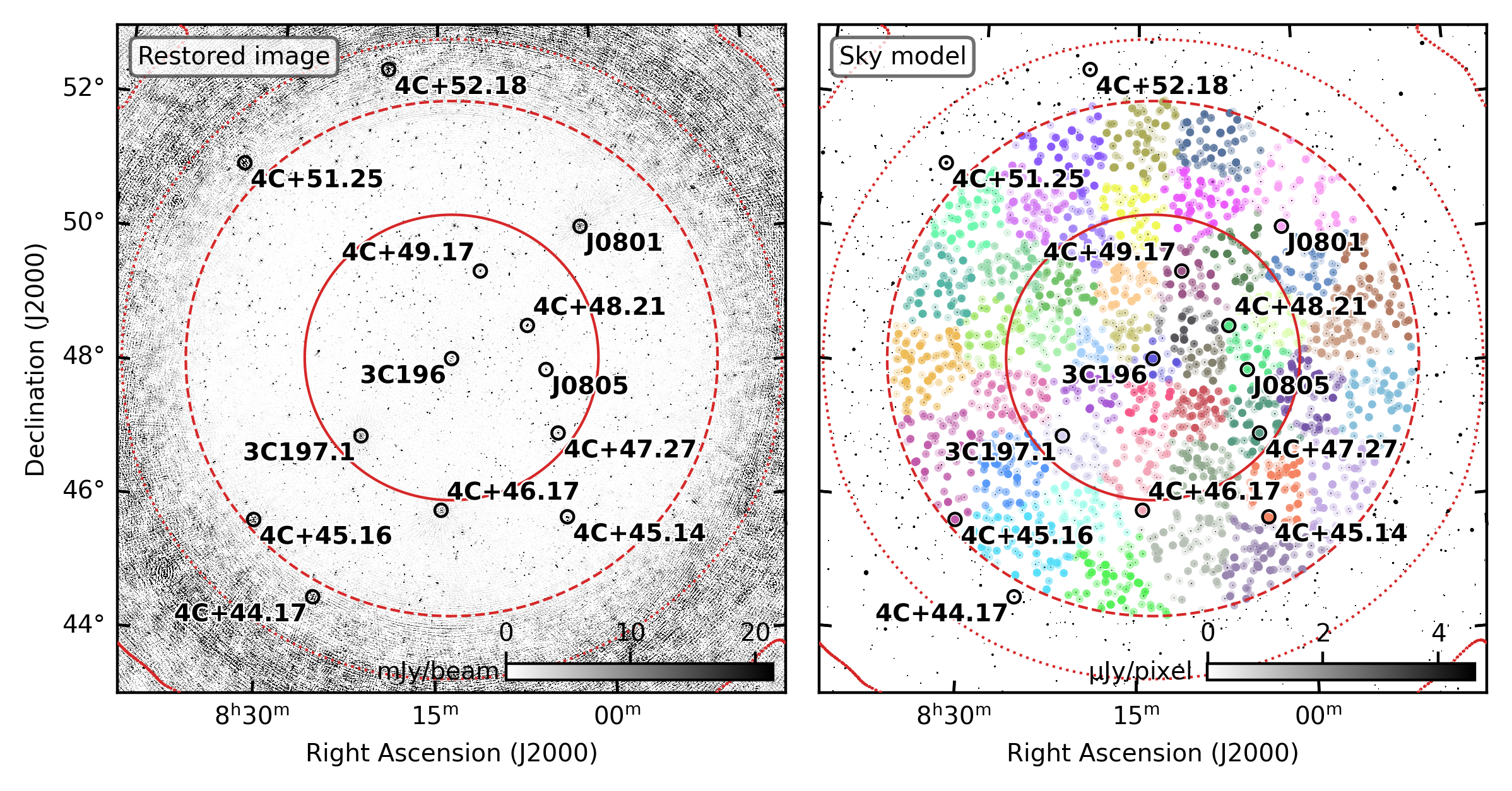}
    \caption{The left panel shows the frequency-integrated (120--168\,MHz) restored image of the 3C\,196 field, where 3C\,196 has been subtracted and the full primary beam correction applied. The image noise is $\sigma = 0.8\,\text{mJy/beam}$. The model image resulting from the imaging process is shown in the right panel, where the sources of the 47 clusters are highlighted by different colours. Both images have a pixel resolution of 3\,arcsec and a field of view of $10^\circ\times10^\circ$. The 1, 7.5, and 50 per cent levels of the primary beam intensity are shown with the red contours. The 7.5 per cent level roughly corresponds to the $3.9^\circ$ radius that we used to select the sky model components. The brightest 3C and 4C sources are also indicated by black circles, in addition to J080135.35+500943.9 (shortened to J0801) and J080508+480151 (shortened to J0805) that have approximately 11 and 2\,Jy of total flux, respectively.}
    \label{fig:3C196_skymodel_image_model_clustered}
\end{figure*}

The sky model of the 3C\,196 field was made by taking the multi-scale clean components (Gaussian and point sources) from the calibrated data using the \textsc{wsclean} imager software \citep{offringa_etal:2014}. We used its multi-frequency, joined-channel deconvolution \citep{offringa_smirnov:2017} to accurately model the spectral features of the sources. This method splits the full bandwidth into multiple channels, while peak finding of the clean components is performed on a frequency-integrated image, which has a higher signal-to-noise ratio. Cleaning is then performed in each output channel at the identified component positions, and spectral smoothness can be enforced by fitting an ordinary or logarithmic polynomial function. In our case, we used 12 output channels, each 4\,MHz wide, and fitted an ordinary, third-order polynomial function with four terms. Although sources at low frequencies are primarily dominated by synchrotron emission and thus have a power-law spectral energy distribution, using a logarithmic function for fitting often results in incorrect spectral index values due to limited bandwidth or significant systematics \citep[e.g.][]{offringa_etal:2016}. While ordinary polynomial functions are generally more stable, they are challenging to use at different frequencies. However, since the sky model of the 3C\,196 field will be used in the LOFAR-EoR pipeline within the frequency range from which it was extracted, this is not an issue. 

To ensure the best quality for each output channel, we did not apply multi-frequency weighting\footnote{More details in \url{https://wsclean.readthedocs.io/en/latest/mf_weighting.html}.}. We used the W-gridder algorithm \citep{arras_etal:2021} for gridding the visibilities. Each output channel was then weighted with a Briggs weighting scheme with a robust parameter of $-1$, resulting in a synthesised beam with a full width at half maximum (FWHM) of $13\times18$\,arcsec at the lowest frequency channel (centred on 122\,MHz). To achieve a more circular beam, we applied a Gaussian taper of 20\,arcsec in size to the gridded $u\varv$-plane. Using a pixel size of 3\,arcsec was sufficient to sample the resulting synthesised beam.

Alongside the multi-frequency deconvolution, we also used the multi-scale \textsc{clean} algorithm \citep{offringa_smirnov:2017} to accurately model the spatial structures of the sources. The multi-scale deconvolution is useful for cleaning resolved sources, representing source models as a summation of basis functions (Gaussian or tapered quadratic) of different sizes and point-like components. Since we are not focusing on a single source and our resolution is degraded by the Gaussian taper, we limited the multi-scale algorithm to four spatial scales, namely 0 (delta function), 1.4, 2.7, and 5.4\,arcmin. The cleaning was performed on the flat-noise images (i.e.\ before the spatially varying primary beam correction) to an initial threshold of $5\sigma$, at which a pixel mask was made by using the auto-masking feature of \textsc{wsclean}, finally cleaning to $1\sigma$ within that mask, where $\sigma = 0.15\,\text{mJy\,beam}^{-1}$ for the frequency-integrated image.

The spatially varying beam is then corrected from the deconvolved images and the clean model to obtain the intrinsic fluxes. The primary beam size and shape vary with frequency, having a narrower main lobe at higher frequencies. Sources that are near or in the predicted beam null can cause stability issues. For these reasons, we focused on imaging mainly the main lobe, which has a FWHM of approximately $4^\circ$ at 150\,MHz, resulting in an image size of $12\text{k} \times 12\text{k}$ pixels to cover a sky area of $10^\circ \times 10^\circ$.

The frequency-integrated continuum image of the 3C\,196 field, with 3C\,196 subtracted and after the full primary beam correction, is shown in the left panel of Fig.~\ref{fig:3C196_skymodel_image_model_clustered}. The red contours indicate the primary beam intensity as a percentage of the peak (which is one at the phase centre) for the frequency-integrated data. This means that, for instance, the 1 per cent level corresponds to a larger area at low frequencies and a smaller area at higher frequencies. In the image, outside the 7.5 per cent contour, the noise boost due to the beam correction becomes appreciable. The cleaning process also found components in this area, as shown in the right panel of Fig.~\ref{fig:3C196_skymodel_image_model_clustered}. Cutting the image field of view at $10^\circ$ was insufficient to exclude such sources, which can cause stability issues due to the low value of the beam response. In the next section, we describe the selection performed on such a raw clean model to obtain the final, cleaned sky model that is used in the 21-cm signal processing pipeline.

\begin{figure*}
    \centering
    \includegraphics[width=0.99\textwidth,trim={5.70cm 0 5.70cm 0},clip]{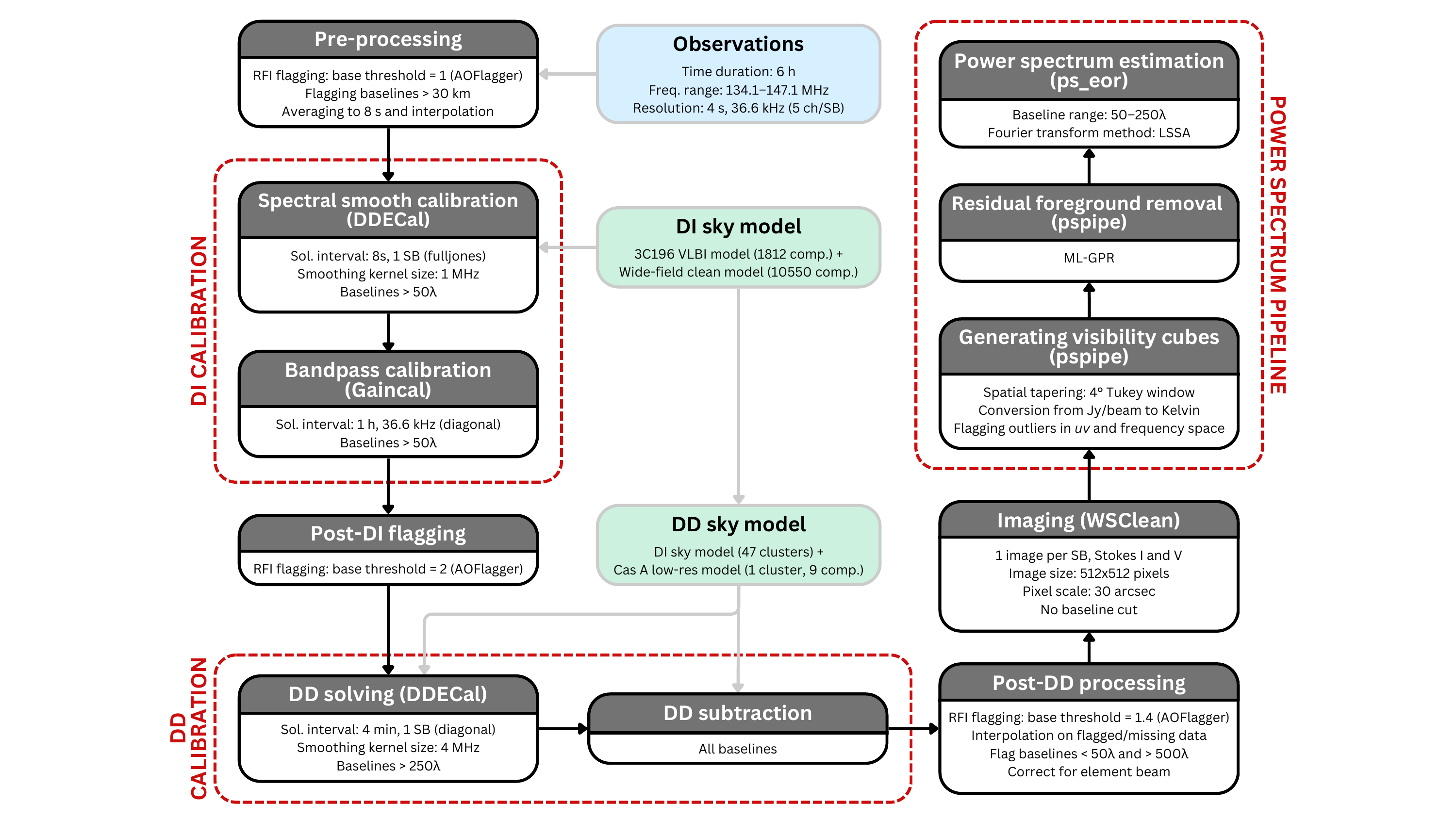}
    \caption{The 21-cm signal processing pipeline for the LOFAR 3C\,196 field, from the pre-processing and DI calibration to the final power spectrum estimation.}
    \label{fig:3C196_pipeline}
\end{figure*}

\medskip

\subsection{Sky model extraction and clustering}\label{sec:mod:sky-model-extraction}

Selection and cleaning of the sky model sources were performed in three steps:
\begin{enumerate}
    \item We removed deconvolved residuals, which are visible in the position of 3C\,196, visible in the restored image on the left panel of Fig.~\ref{fig:3C196_skymodel_image_model_clustered}. This was done by cutting out all the cleaned components within a central aperture with a radius of 2\,arcmin.
    
    \item We only kept components within an aperture centred on 3C\,196 with a radius of $3.9^\circ$, corresponding to the 7.5 per cent level of the frequency-integrated primary beam main lobe, shown in Fig.~\ref{fig:3C196_skymodel_image_model_clustered}. The components selected in this way were not affected by the weaker response of the primary beam.
    
    \item We removed deconvolution artefacts that resulted in isolated negative components, although the total intensity of these components was only $-231\,\text{mJy}$ at 150\,MHz.
\end{enumerate}
These operations resulted in a final sky model with 10\,550 components, to which we added the 3C\,196 model used in the DI calibration, bringing the total to 12\,362 components. The resulting model is referred to as the `DI sky model' because it is used for the DI calibration steps of the 21-cm signal processing pipeline.

This model is then split into multiple directions (or clusters) in the sky for direction-dependent (DD) calibration. Clustering was done using a modified $K$-means algorithm that considers angular distances instead of Euclidean ones. Therefore, due to projection effects, outer clusters are more extended than central ones. The $K$-means algorithm does not account for the flux density of clustered sources, so we validated that each direction had sufficient signal-to-noise ratio during DD solving. We aimed for a flux density greater than $5\sigma_\text{DD}$, where $\sigma_\text{DD}$ is the noise level within a DD calibration gain interval, calculated using the radiometer equation:
\begin{equation}
\sigma = \frac{\text{SEFD}}{\sqrt{2\Delta t \Delta \nu}}\, ,
\end{equation}
where SEFD is the system equivalent flux density, $\Delta t$ is the time interval, and $\Delta \nu$ is the frequency interval. For DD gain solutions (described in Section~\ref{sec:proc:dd-cal}), we used $\Delta t = 4\,\text{min}$ and a spectral smoothing kernel of 4\,MHz, approximating $\Delta \nu$ with this value. With an average SEFD of 2835\,Jy per visibility for our data set, we get $\sigma_\text{DD}\approx 65\,\text{mJy}$, meaning each cluster needs at least 325\,mJy.

We reached this flux level with our sky model by using 60 directions, but this resulted in small clusters containing only a few tens of components.  Clusters should generally be $1{-2}^\circ$ in size to ensure enough signal-to-noise for the solutions, while also reducing the spatially varying DD effects. In addition, having too many clusters also increases the degrees of freedom in the calibration, which raises both computational cost and the risk of absorbing unmodelled signal into the solutions. To address both the signal-to-noise and degree-of-freedom aspects, we aimed to ensure that each cluster contained more than 100 components and was approximately $1^\circ$ in size, opting for 47 directions. This resulted in all clusters having more than 1\,Jy total flux and more than 110 components (see Appendix~\ref{app:dd-skymodel}). Components within the same cluster are highlighted in the same colour in the right panel of Fig.~\ref{fig:3C196_skymodel_image_model_clustered}. To this model, we finally add a low-resolution model of Cas\,A\footnote{\url{https://github.com/lofar-astron/prefactor/blob/master/skymodels/A-Team_lowres.skymodel}} with nine components in a separate cluster. The resulting model is referred to as the `DD sky model' from here onward.

\section{3C\texorpdfstring{\,}{}196 processing pipeline}\label{sec:3c196-processing}

The data processing for the 3C\,196 field is based on an updated pipeline from the NCP processing \citep{mertens_etal:2020}, where we replaced \textsc{sagecal} by  \dppp. This choice was motivated by the need for a well-integrated environment that provides tools for processing, flagging, and both DI and DD calibration, while also supporting sky models in the format output by \textsc{wsclean}. These tools are the standard for processing LOFAR observations and are implemented in pipelines such as \textsc{linc}\footnote{\url{https://linc.readthedocs.io/en/latest/index.html}} \citep[LOFAR initial calibration;][]{degasperin_etal:2019} and \textsc{rapthor}\footnote{\url{https://rapthor.readthedocs.io}}. A fully \dppp-based 21-cm signal processing pipeline has been developed and used by \cite{munshi_etal:2024:ul} to set the first upper limits on the Cosmic Dawn signal from NenuFAR.

The 3C\,196 processing pipeline is designed to account for both DI and DD effects while preserving the cosmological 21-cm signal. It consists of several stages: (1) pre-processing and RFI flagging, (2) DI calibration, (3) DD solving and sky model subtraction, (4) imaging, (5) modelling and removal of residual foregrounds, and (6) power spectrum estimation. The pipeline is applied to the frequency range $134.2{-}147.1\,\text{MHz}$, corresponding to the redshift bin centred at $z=9.16$. Steps (1) and (2) are similar to the processing for generating the sky model (Section~\ref{sec:sky-modelling}). In the following, we provide a detailed description of each step of the 3C\,196 21-cm signal pipeline, an overview of which is presented in Fig.~\ref{fig:3C196_pipeline}. Step~(5) involves applying the GPR method \citep{mertens_etal:2018} and is described in Section~\ref{sec:gpr}.

\subsection{Pre-processing}\label{sec:proc:pre-proc}

The initial data set had already undergone partial pre-processing when the observations were taken in 2014, including averaging to 5 channels per SB and 4\,s time intervals. However, we found residual RFI that had not been properly removed. To address this, we used \textsc{aoflagger} with an updated RFI flagging strategy optimised for LOFAR-HBA data. \textsc{aoflagger} uses an adaptive thresholding algorithm to automatically detect and flag corrupted data. During the initial pre-processing, we set the base threshold to the default value of 1, which corresponds to a flagger false-positives rate of 0.5 per cent \citep{offringa_etal:2013}\footnote{For more details about the algorithm parameters, we refer to \citet{offringa_etal:2010} and \url{https://aoflagger.readthedocs.io}}. This resulted in about 3.4 per cent of visibilities flagged. Additionally, we removed baselines longer than 30\,km using the same method as described in Section~\ref{sec:mod:baselines} to minimize the effects of ionospheric distortions.

Following the RFI and baseline flagging, the data were then averaged to 8\,s time intervals. This averaging step reduces the data volume, thereby decreasing computational costs in subsequent processing steps, without decorrelating the signal from the longer baselines. To further mitigate the impact of flagged or missing data, a Gaussian-weighted interpolation scheme was applied. This interpolation ensures that the introduced values are smooth and consistent with the surrounding data, preventing the generation of artificial spectral fluctuations (see \citealt{offringa_etal:2019a} for details).

\subsection{Direction-independent calibration}\label{sec:proc:di-cal}

The DI calibration is performed similarly to Section~\ref{sec:mod:di-cal}, solving first for spectrally smooth gains on short time intervals and then for the high spectral resolution bandpass response on long time intervals. We used the DI sky model for both steps. Because the model is intrinsic by construction, we had to apply the full LOFAR primary beam model. For the spectrally smooth calibration, we solved for the high-temporal effects, using the \textsc{ddecal} tool to apply a Gaussian smoothing kernel of 1\,MHz width on the per-station solutions at every iteration. We solved for full Jones matrices with a time interval of 8\,s (i.e.\ one integration time) and a frequency interval of 183.1\,kHz (i.e.\ one SB). The resulting solutions are shown in Fig.~\ref{fig:cal_smooth}. Both amplitudes and phases were applied to the data.

After the spectral smooth calibration, a bandpass calibration was performed using the \textsc{gaincal} tool to correct for the frequency response of the instrument. We solved for diagonal gains to reduce the number of free parameters, using a time interval of 1\,h and a frequency interval of 36.6\,kHz (i.e.\ one channel). The gain solutions are shown in Fig.~\ref{fig:cal_bp}. As with the spectral smooth calibration, we applied both amplitudes and phases to the data. 

A second round of RFI flagging was performed on the DI calibrated data with a decreased sensitivity (an \textsc{aoflagger} base threshold of 2) compared to the pre-calibration flagging. This step removed residual RFI that may have been left unflagged during the initial pre-processing, without strongly affecting the data by using a higher threshold. Only 0.9 per cent of the remaining data were flagged. 

\begin{figure*}
    \centering
    \includegraphics[width=0.95\textwidth]{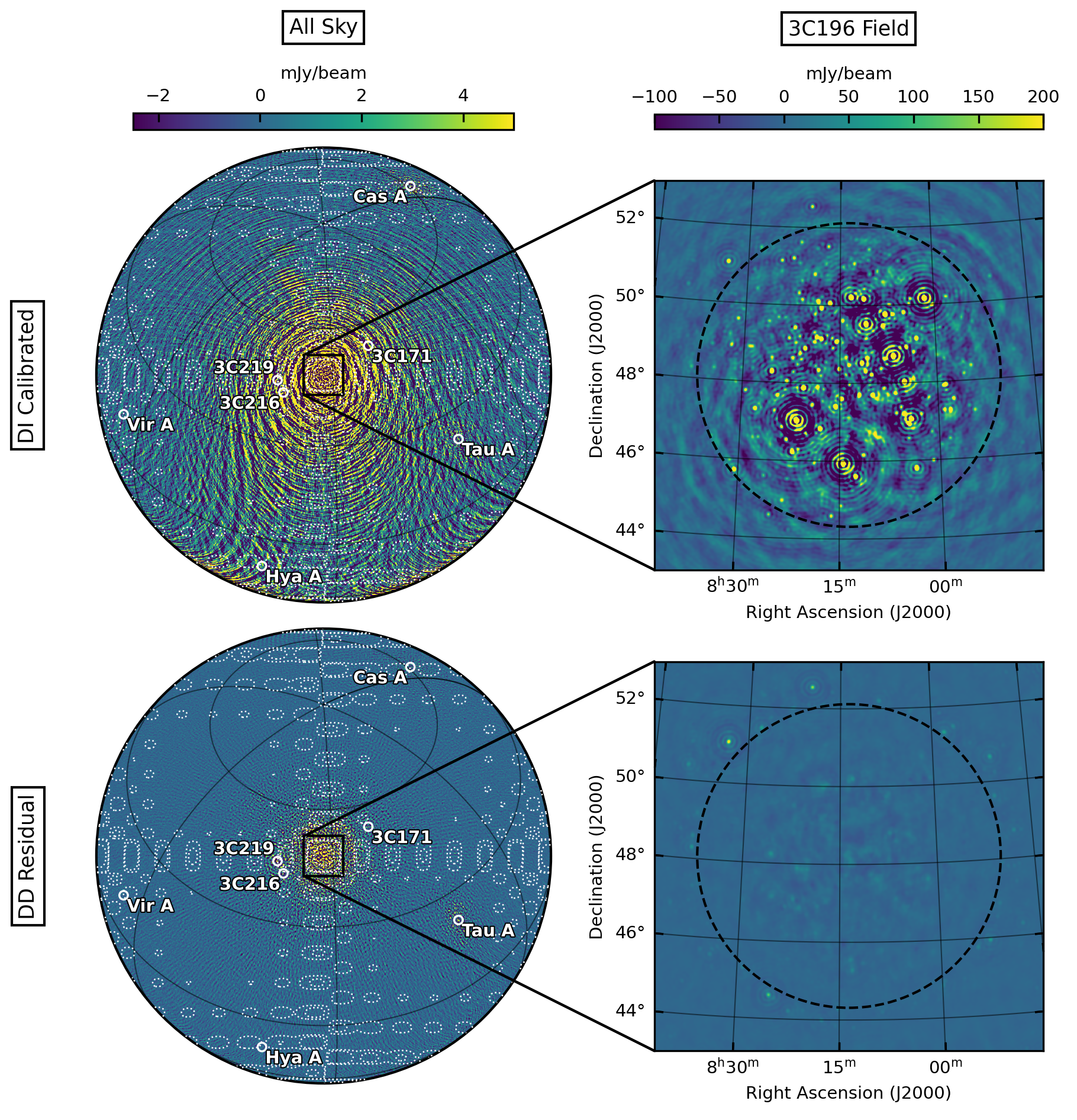}
    \caption{Frequency-integrated (134--147\,MHz) dirty images of the 3C\,196 field after the DI calibration with only 3C\,196 subtracted (top row) and after the DD subtraction (bottom row). The left column shows the all sky images with a pixel scale of 5\,arcmin, where a Gaussian taper with $\text{FWHM}=10\,\text{arcmin}$ and a Briggs weighting with robust parameter $-2$ have been applied to the $u\varv$-coverage to better highlight the bright sources. The same colour range has been used in the top and bottom panels to show the level of source sidelobe suppression after the DD subtraction step. The 1, 5, and 50 per cent levels of the time and frequency averaged primary beam are plotted with the dotted contours. The brightest sources are indicated with white circles, including the A-team sources. The right column reports the zoom-in of the 3C\,196 central field, imaged with natural weighting and a pixel scale of 0.5\,arcmin, covering a $10^\circ \times 10^\circ$ field of view. Only the baselines between 50 and 500$\lambda$ have been selected. Also here, the same colour range is used in the top and bottom panels. The dashed black circle highlight the $3.9^\circ$ radius extension of our sky model.}
    \label{fig:3C196_DI_DD_widefield_zoomin}
\end{figure*}

\begin{figure}
    \centering
    \includegraphics[width=1\columnwidth]{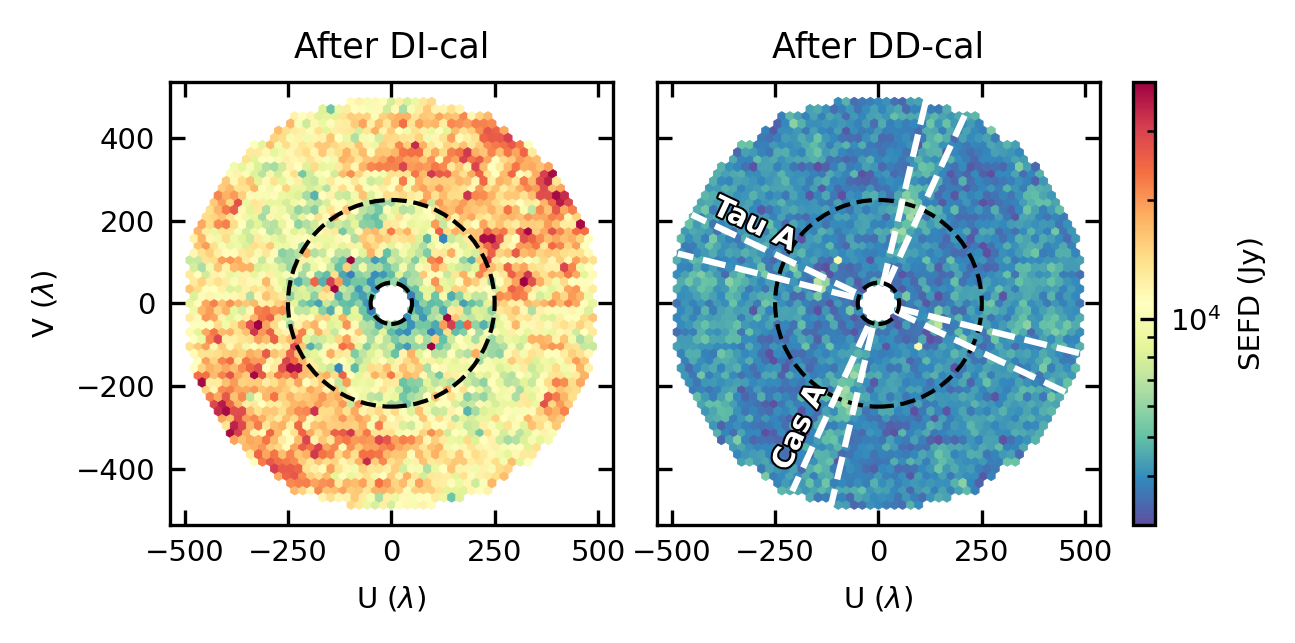}
    \caption{Standard deviation of the Stokes~I data after DI calibration (left) and DD subtraction (right), expressed in SEFD units, where the SEFD is calculated from the frequency-differenced thermal noise at each $u\varv$-cell in the $50{-}500\lambda$ range. The inner and outer dashed black circles indicate the 50 and 250$\lambda$ cut used for the power spectrum estimation. The white dashed lines represent the range where we expect the sidelobe contributions of Cas\,A and Tau\,A to dominate, given their direction in the sky with respect to the phase centre.}
    \label{fig:UVplane_DI_DD}
\end{figure}

The dirty images of the DI calibrated data, made with \textsc{wsclean}, are shown in the top panels of Fig.~\ref{fig:3C196_DI_DD_widefield_zoomin}, with the all-sky image shown on the left, while a zoom-in of the central 3C\,196 field is shown on the right, with the same size of Fig.~\ref{fig:3C196_skymodel_image_model_clustered}. Both the continuum images are frequency-integrated within the redshift bin bandwidth (134.2--147.1\,MHz). We subtracted 3C\,196 to make the rest of the field visible. A Briggs weighting scheme with robust parameter of $-2$ and a $u\varv$-coverage Gaussian taper of 10\,arcmin have been used for the all-sky image, while natural weighting and $u\varv$-range of $50{-}500\lambda$ have been set for the zoom-in. The all-sky image is dominated by sidelobes from the central field sources, which are still visible despite the use of the Gaussian taper, making even bright A-team sources hard to see. No point spread function (PSF) sidelobes of A-team sources are evident in the zoomed-in image, which is instead dominated by 3C and 4C sources. In fact, the 3C\,196 field presents many more of these sources than the NCP, with flux densities higher than 1\,Jy, both inside and outside the main primary beam lobe. We also calculated the standard deviation of the Stokes~I gridded data for each $u\varv$-cell. This is shown in the left panel of Fig.~\ref{fig:UVplane_DI_DD} in SEFD units, between 50 and $500\lambda$. The power seems uniformly spread over the $u\varv$-grid, with no clear directionality visible, which suggests that it comes from many sources \citep{munshi_etal:2025}, unlike the NCP field where Cas\,A and Cyg\,A dominate the sidelobe leakage into the central field \citep{mertens_etal:2018,gan_etal:2022}.

\subsection{Direction-dependent calibration}\label{sec:proc:dd-cal}

\begin{figure}
    \centering
    \includegraphics[width=1\columnwidth]{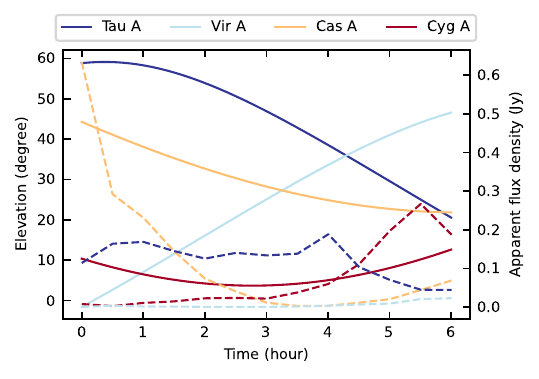}
    \caption{Elevations (left axis, solid lines) and apparent flux densities (right axis, dashed lines) of the brightest A-team sources: Cygnus\,A (Cyg\,A), Cassiopeia\,A (Cas\,A), Virgo\,A (Vir\,A), and Taurus\,A (Tau\,A). Line colours correspond to the distance from the phase centre, with dark blue representing the nearest source (Tau\,A) and dark red the most distant (Cyg\,A). Apparent flux densities were calculated at 30-minute intervals by multiplying the total flux density of each source by a time and frequency-averaged primary beam model value in the source direction.}
    \label{fig:ateam_elevation_appflux}
\end{figure}

The large field of view of LOFAR introduces significant DD effects due to the time-varying ionosphere and the imperfect knowledge of the primary beam. These effects are addressed by performing a DD calibration, where the sky model is divided into multiple directions (or clusters), and solutions are derived for each direction. Because of the bright sources present in the 3C\,196 field, clusters of $1^\circ$ were large enough to fulfil these conditions, resulting in a DD sky model with 47 directions as described in Section~\ref{sec:mod:sky-model-extraction}. In addition, we included a model of Cas\,A, which is the brightest A-team source in apparent flux, as shown with dashed lines in Fig.~\ref{fig:ateam_elevation_appflux}, but only during the first quarter of the observation. The source is at the edge of a primary beam sidelobe, plotted with dotted white lines in the left column of Fig.~\ref{fig:3C196_DI_DD_widefield_zoomin}, and its elevation decreases over the night. The second brightest source is Taurus\,A (Tau\,A), which is also the closest to the main field. Because we do not have a well-tested model of it, we do not include Tau\,A in this work. In general, subtracting an incorrect source model can introduce more residual contamination than omitting it entirely. Instead, we let GPR handle the source sidelobes, which may lead to better results, as shown in the latest NCP analysis, where removing poorly modelled sky clusters improved the results \citep{mertens_etal:2025}. 

\begin{figure*}
    \centering
    \includegraphics[width=0.95\textwidth]{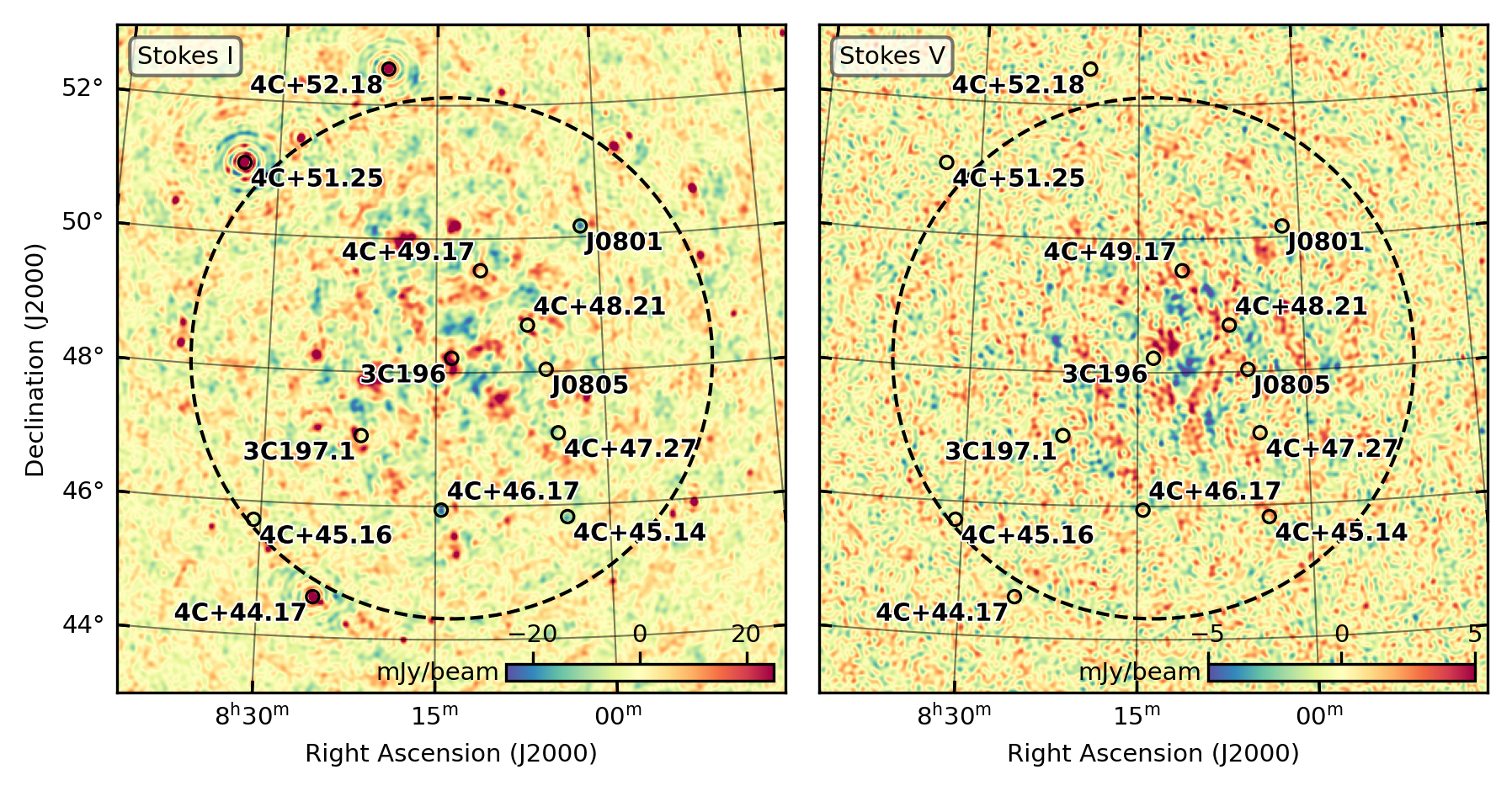}
    \caption{Restored Stokes~I (left) and Stokes~V (right) images of the SB centred at 140.4\,MHz. Both images have a pixel scale of 0.5\,arcmin and a field of view of $10^\circ \times 10^\circ$. The $u\varv$-plane was gridded with a natural weighting scheme and only baselines between 50 and $500\lambda$ were used. The black dashed circle indicates the $3.9^\circ$ radius of the sky model. The brightest sources are highlighted with black circles.}
    \label{fig:3C196_afterDD_StokesI_V}
\end{figure*}

The DD calibration consists of two steps: (i) solving for the frequency-dependent gain solutions per source cluster, and (ii) subtraction of the sky model `corrupted' by these gains. Thus, the gain solutions are not applied to the data but to the model. To find the DD solutions, we used the directional solving algorithm of \textsc{ddecal} \citep{smirnov_tasse:2015}, which allowed us to apply a frequency smoothing kernel to the gains, similarly to the spectral smooth DI calibration. In this case, we used a Gaussian kernel with a 4\,MHz width, which yielded better results than the 1\,MHz width for the NCP field \citep{gan_etal:2023}. The gains are solved iteratively for each direction and diagonal elements of the Jones matrix, setting 4\,min and 183.1\,kHz (i.e.\ one SB) as solution time and frequency intervals, respectively. This yields enough signal-to-noise for each direction, as discussed in Section~\ref{sec:mod:sky-model-extraction}. For the NCP a shorter time interval (i.e.\ 2.5\,min) had to be selected for far-field sources, such as Cas\,A and Cyg\,A, because no primary beam was applied and shorter time-scale fluctuations in the beam needed to be solved for. By using an intrinsic sky model and applying the primary beam, we can keep the same time interval for both in and far field sources. Similarly to the NCP data processing, we excluded baselines shorter than $250\lambda$ while solving to avoid over-fitting and signal loss in the $u\varv$ range of interest for the 21-cm signal extraction, which is $50{-}250\lambda$ \citep{mouri_sardarabadi_etal:2019,mertens_etal:2020,mevius_etal:2022}. The solutions are shown in Fig.~\ref{fig:cal_dd} for the main field clusters, and in Fig.~\ref{fig:cal_dd_casa} for the Cas\,A direction. 

After subtracting the sky model with the DD gains applied, a final round of RFI flagging was performed using a base threshold of $1.4$, which we found sufficient to remove residual low-level RFI. This resulted in only 0.13 per cent of additional flagged data. We also interpolated the flagged and missing data to reduce the excess noise caused by artificial spectral fluctuations \citep{offringa_etal:2019a}. During this post-DD processing, we flagged baselines shorter than $50\lambda$ and longer than $500\lambda$. This $u\varv$-cut is usually applied only during the next imaging step \citep{mertens_etal:2020}, but in that case it is performed on the gridded $u\varv$-plane, and hence it depends on the image size. Larger images mean smaller $u\varv$-cells, while smaller images mean $u\varv$-cells spanning a larger interval of real $u\varv$ lengths. Cutting at $50\lambda$ in the gridded $u\varv$-plane would also result in excluding baselines longer than $50\lambda$, biasing the power spectra by setting the image size. A better approach is to directly flag the excluded baselines after DD subtraction and use no $u\varv$-cut in the imaging step. Finally, we corrected for the LOFAR element beam. This correction is not necessary in the NCP pipeline because the NCP sky model already has this scaling factor embedded, and the DD gain solutions take it into account. Thus, the power spectrum pipeline was designed to correct only for the array factor. In our case, applying the full primary beam both during gain solving and application kept the visibility data fully apparent, and therefore required the element beam correction to get correct power spectrum values. 

The results after the DD subtraction and final processing are shown in the bottom panels of Fig.~\ref{fig:3C196_DI_DD_widefield_zoomin}, where in the all-sky dirty image (bottom left panel) we see that the bright sidelobes from the main field sources have been strongly suppressed and allows us to better distinguish far-field sources such as Tau\,A. With the same colour scale as for the DI calibrated data (top right panel), the DD subtraction residuals (bottom right panel) show how well the bright sources in the central $3.9^\circ$ radius (black dashed circle) have been removed. The noise standard deviation of the frequency-integrated images went from $\sigma = 13.6\,\text{mJy/beam}$ of the DI calibrated image to $\sigma = 1.7\,\text{mJy/beam}$ of the residuals after the sky model subtraction. The standard deviation of the $u\varv$-gridded Stokes~I data after the DD subtraction still shows some residual emission both within and outside $250\lambda$, as shown in the right panel of Fig.~\ref{fig:UVplane_DI_DD}. However, a reduction of more than one order of magnitude happened, which allows us to see that part of the residual emission comes from Cas\,A and Tau\,A, besides a few bright $u\varv$-cells where the contamination might come from sources closer to the primary beam main lobe. When the colour scale range of the Stokes~I residuals is reduced as shown in Fig.~\ref{fig:3C196_afterDD_StokesI_V}, we see that there are a few bright sources not included in the DD sky model but very close to the $3.9^\circ$ radius cut, the sidelobes of which do affect the central field. Moreover, some extended emission is visible in the Stokes~I residuals, which might be un-modelled sky emission or due to calibration errors during the DD solving.

The right panel of Fig.~\ref{fig:3C196_afterDD_StokesI_V} shows the Stokes~V image of the SB centred at 140.4\,MHz. The Stokes~V parameter characterises the circular polarization, which is expected to be close to zero for most of radio sources, and can be used to estimate the noise level. However, we found some deviation from the noise level, especially close to the centre of the field. We do not have a sky model for parameters other than Stokes~I, and our calibration did not therefore correct for all the polarization leakage. \citet{jelic_etal:2015} showed that the 3C\,196 field exhibits strong polarized structures, which are somewhat similar to what we observed in our Stokes~Q and U images (see Appendix~\ref{app:polQU}). Leakage from Stokes~U into Stokes~V is possible, as both are calculated from the XY and YX elements of visibilities. However, we found that the structures in Stokes~V more closely resemble those observed in Stokes~Q.

\subsection{Imaging and conversion to Kelvin}\label{sec:proc:imaging}

After the DD calibration and final processing steps, the residual visibilities are gridded and imaged using \textsc{wsclean}, creating $(l,m,\nu)$ image cubes. To minimize aliasing effects during the gridding, we used a Kaiser-Bessel filter with a kernel size of 15 $u\varv$-cells, an oversampling factor of 4096, and 32 $w$-layers, similar to the settings by \citet{mertens_etal:2020}. \citet{offringa_etal:2019b} showed that these settings keep artefacts introduced by the gridding well below the level of the cosmological 21-cm signal.

The imaging was performed separately for each SB, allowing the production of even and odd time-step Stokes~I and V images. The time-differenced Stokes~V images will be used to estimate the thermal noise variance. For this reason, having some polarization leakage into Stokes~V is not an issue as long as the contamination is not varying rapidly over time. We used a natural weighting scheme, so that every visibility was given a constant weight. We did not set any baseline cut because of the previous baseline flagging, where we selected only the $50{-}500\lambda$ range. We set a pixel scale of 30\,arcsec and an image size of $512 \times 512$ pixels, which covers a field of view of approximately $4.3 \times 4.3^\circ$.

To convert the images from units of Jy/beam to units of brightness temperature (Kelvin), we Fourier transformed the $(l,m,\nu)$ image cube into a gridded $(u,\varv,\nu)$ visibility cube. The conversion was then performed using the method outlined by \citet{offringa_etal:2019b}. A spatial Tukey taper with a $4^\circ$ diameter was also applied to focus on the central part of the primary beam, which has $\text{FWHM}\approx4.1^\circ$ at 140\,MHz, and to avoid sharp image edges.

\begin{figure}
    \centering
    \includegraphics[width=1\columnwidth]{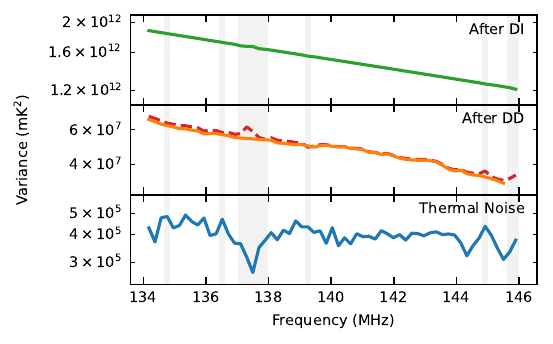}
    \caption{Variance in $\text{mK}^2$ units as a function of frequency for the Stokes~I after DI calibration (green solid line), after DD subtraction (red dashed line), and after final outlier flagging (orange solid line, see Section~\ref{sec:proc:imaging}). The thermal noise estimated from the time-differenced Stokes~V visibilities is shown with the blue solid line. The gray shaded areas represent the flagged channels.}
    \label{fig:Variance}
\end{figure}

Before proceeding to power spectrum estimation, we performed additional flagging to remove any outliers in the gridded data cubes, using a $k$-sigma clipping method with de-trending. These outliers are mainly low-level RFI that \textsc{aoflagger} failed to detect. In particular, $u\varv$-grid outliers are flagged based on their weights (i.e.\ the per-visibility inverse variance combined with the weights given by the $u\varv$-density of the gridded visibilities), Stokes~V variance, standard deviation of time-differenced Stokes~V, and standard deviation of frequency-differenced Stokes~I (i.e.\ right panel of Fig.~\ref{fig:UVplane_DI_DD}). Frequency outliers are flagged based on their weights and standard deviation of frequency-differenced Stokes~V. Figure~\ref{fig:Variance} shows the variance in $\text{mK}^2$ over frequency for the Stokes~I residual cubes before (red dashed line) and after (orange solid line) this last flagging. The shaded grey areas represent the resulting flagged channels. We found that 6.4 per cent of $u\varv$-cells and 16.4 per cent of the SB are flagged, which are percentages lower than the values found in the NCP processing \citep{mertens_etal:2020}.

\subsection{Power spectrum estimation}

Let $T(\mathbf{x})$ represent the brightness temperature of a signal at a physical coordinate $\mathbf{x}$. The corresponding power spectrum, denoted as $P(\mathbf{k})$, is a function of the wavenumber $\mathbf{k}$ (with units of $h\,\text{cMpc}^{-1}$) and can be expressed as
\begin{equation}\label{eq:PS}
    P(\mathbf{k}) = V | \tilde{T}(\mathbf{k}) |^2 \, ,
\end{equation}
where $V$ is the observing comoving volume, and $\tilde{T}$ is the discrete Fourier transform of the brightness temperature. This power spectrum is typically reported in units of $\text{K}^2\,h^{-3}\,\text{cMpc}^3$. The components of $\mathbf{k}$, perpendicular and parallel to the line of sight, are respectively given by \citep{morales_hewitt:2004,vedantham_etal:2012}:
\begin{equation}\label{eq:kperp_kpar}
    \mathbf{k}_\perp = \frac{2\pi}{D_M(z)} \mathbf{u}\, ,\ k_\| = \frac{2\pi H_0 \nu_{21} E(z)}{c (1 + z)^2}\, \eta\, ,
\end{equation}
where $D_M(z)$ is the transverse comoving distance at redshift $z$, $\nu_{21} = 1420\,\text{MHz}$ is the rest frequency of the neutral hydrogen hyperfine transition line, $H_0$ is the Hubble constant, $E(z)$ represents the dimensionless Hubble parameter, $c$ is the speed of light, and $\eta$ is the Fourier dual of the frequency $\nu$.

The cylindrically averaged (2D) power spectrum can be obtained by averaging $P(\mathbf{k})$ in cylindrical shells with radius $k_\perp = |\mathbf{k}_\perp|$:
\begin{equation}\label{eq:PS2D}
    P(k_\perp, k_\|) = \frac{\sum_{\mathbf{k} \in (k_\perp, k_\|)} P(\mathbf{k})}{N_{(k_\perp, k_\|)}} \, ,
\end{equation}
where $N_{(k_\perp, k_\|)}$ is the number of $\mathbf{k}$-space cells within the $(k_\perp, k_\parallel)$-annulus. Alternatively, the dimensionless spherically averaged (1D) power spectrum can be derived by averaging in spherical shells with radius $k = |\mathbf{k}| = \sqrt{k_\perp^2 + k_\|^2}$:
\begin{equation}\label{eq:PS1D}
    \Delta^2(k) = \frac{k^3}{2\pi^2} \frac{\sum_{\mathbf{k} \in k} P(\mathbf{k})}{N_k} \, ,
\end{equation}
where $N_k$ is the number of $\mathbf{k}$-space cells within the $k$-shell. In 21-cm cosmology, the spherical power spectrum is in units of $\text{K}^2$, and because of the direct connection to brightness temperature units, the 21-cm signal upper limits are typically reported as $\Delta^2(k)$. In Equations~\eqref{eq:PS2D} and \eqref{eq:PS1D}, $P(k_\perp, k_\|)$ and $\Delta^2(k)$ are optimally weighted according to the gridded visibility thermal noise \citep[more details in][]{mertens_etal:2020}.

The wavenumber $k_\|$ is effectively the Fourier conjugate of frequency, and spectrally smooth foregrounds should remain confined at low $k_\|$ values, whereas the 21-cm signal affects a wider range of $k_\|$ modes \citep[e.g.][]{santos_etal:2005}. In addition to this mode separation, we have to consider the intrinsic chromaticity of interferometers, which spreads foreground power at higher $k_\|$, causing a process called `mode-mixing' \citep{morales_etal:2012,morales_etal:2019}. Because of the characteristic shape that this effect assumes in the cylindrically averaged power spectrum, it is also denoted as the `foreground wedge' \citep{datta_etal:2010,vedantham_etal:2012,liu_etal:2014a,liu_etal:2014b}. The extent of the foreground wedge is defined by the horizon delay line, which can be derived under a formalism accounting for curved sky effects following \citet{munshi_etal:2025} for phase-tracking instruments such as LOFAR. The same formalism also prescribes source lines which specify a range in the cylindrical power spectrum where emission from a specific direction in the sky is expected to cause power. Because all the sky emission stays within the wedge under reasonable assumptions, above the horizon line there is a foreground-free region called `EoR window', where ideally the power spectrum should be consistent with the noise power spectrum for the sensitivity of the current generation of interferometers. 

\begin{figure*}
    \centering
    \includegraphics[width=1\textwidth]{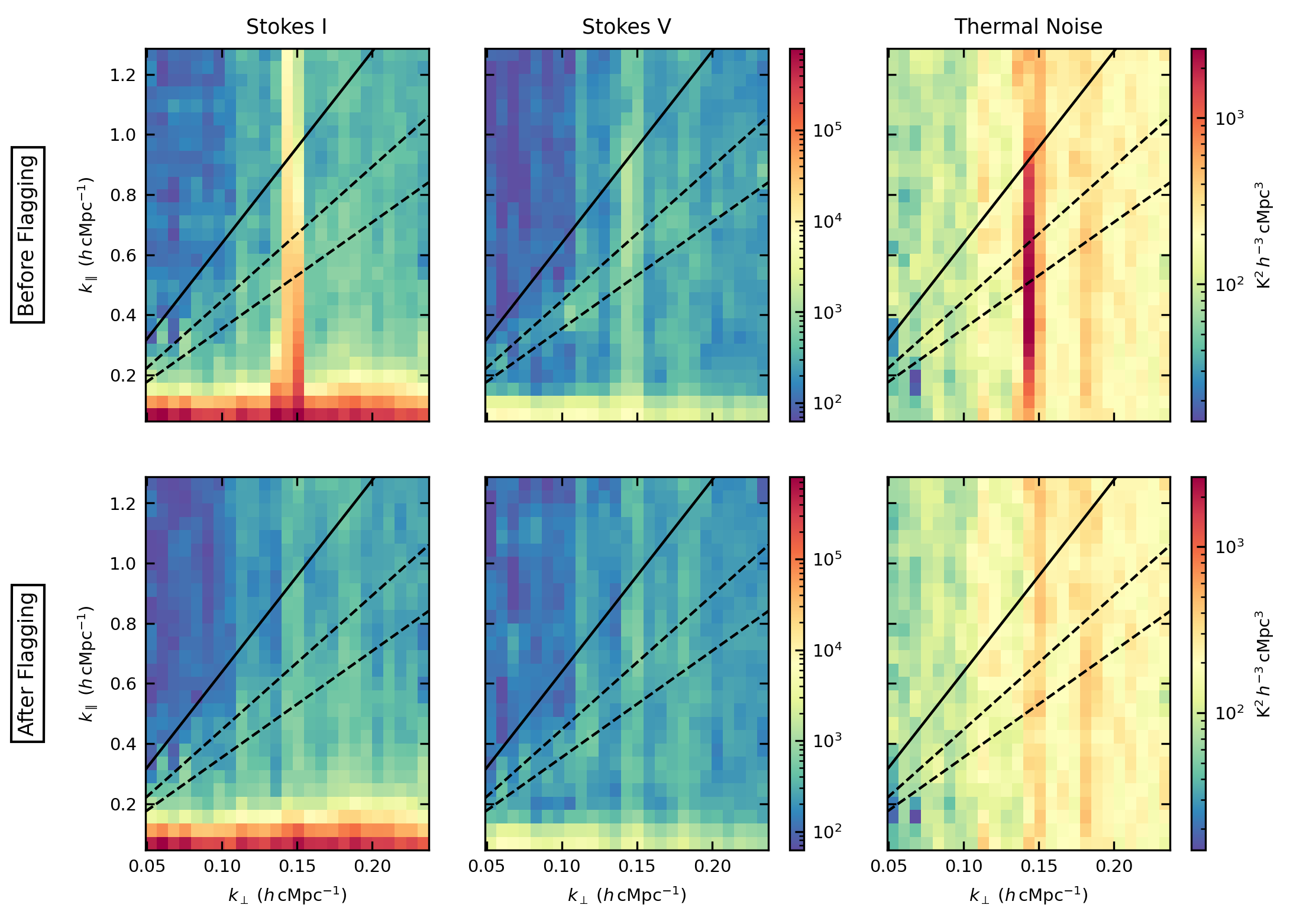}
    \caption{Cylindrical power spectra of Stokes~I (first column), Stokes~V (second column), and thermal noise (third column). The top row shows the power spectra before the outlier flagging discussed in Section~\ref{sec:proc:imaging}, while the bottom row shows them after such a flagging. In all panels, the solid line indicates the horizon delay line, whereas the dashed lines indicate the delay ranges where we expect most of the power of Cas\,A. The same colour bar is used for the panels of the same column, and for Stokes~I and Stokes~V power spectra.}
    \label{fig:PS2D_before_after_flag}
\end{figure*}

The power spectra presented in this paper were estimated using the power spectrum pipeline \textsc{pspipe}\footnote{\url{https://gitlab.com/flomertens/pspipe}}. We selected the $50{-}250\lambda$ baseline range, filtering out the short and long baselines. The reported power spectrum uncertainties were estimated from the sample variance, as described by \citet{mertens_etal:2020}. 

In Fig.~\ref{fig:PS2D_before_after_flag}, we show the cylindrical power spectra of Stokes~I, V, and thermal noise (i.e.\ time-differenced Stokes~V) before (top row) and after (bottom row) the outlier flagging discussed in Section~\ref{sec:proc:imaging}. We applied a Blackman-Harris window function to mitigate the effect of the limited frequency bandwidth leaking power into the EoR window. The power spectra before the flagging show a peak of power at $k_\perp \approx 0.14\,h\,\text{cMpc}^{-1}$, corresponding to a gap in the $u\varv$-coverage which has been observed also in NCP data \citep{mertens_etal:2020}. Flagging the outliers in the $u\varv$-plane lowered the variance, making the thermal noise more uniform and the residual foreground confined at low $k_\|$. 
%
%
%
As seen in Fig.~\ref{fig:3C196_afterDD_StokesI_V}, the power spectrum of Stokes~V shows residual emission at low $k_\|$ because of polarization leakage. However, this leakage is time-correlated and is not visible in the thermal noise, making this effect not a concern for the current analysis. Similar to Fig.~\ref{fig:UVplane_DI_DD}, the cylindrical power spectra also show some residual emission in the direction of Cas\,A (dashed black lines), which is mainly visible in Stokes~V and at low $k_\perp$ in Stokes~I. The outlier flagging removed most of the affected $u\varv$-cells, leaving some residuals only at $k_\perp\lesssim 0.08\,h\,\text{cMpc}^{-1}$.

\section{Residual foreground removal}\label{sec:gpr}

As shown in Fig.~\ref{fig:PS2D_before_after_flag}, the cylindrical power spectra are still dominated by residual foreground contamination, especially at low $k_\|$. These modes are important because it is where the 21-cm signal is expected to be stronger. These residuals are dominated by diffuse Galactic emission, sources near or below the confusion noise, and sources outside the $3.9^\circ$ radius that are not included in the DD sky model. The distinction between foregrounds and the 21-cm signal lies in their spectral behaviour: foregrounds exhibit a much larger coherence scale than both thermal noise and the 21-cm signal. By subtracting smoothed solutions during the DD calibration step, we made sure that the residual foreground contamination also remained smooth. This property can be used to isolate and subtract the residual foregrounds. In this work, we used the GPR method described by \citet{mertens_etal:2018} to model these components. The model basis of the 21-cm signal has been built from simulations by using a machine learning (ML) training, which resulted in an enhanced GPR method called ML-GPR \citep{mertens_etal:2024}.

\subsection{Gaussian process regression}

Gaussian process regression is a non-parametric Bayesian method used to model and predict data based on prior knowledge, where the data is assumed to follow a Gaussian process. In the context of 21-cm cosmology, GPR models the observed data as a sum of Gaussian processes that represent the foregrounds, thermal noise, and the 21-cm signal, each characterised by a specific frequency covariance function $\kappa$ (also called kernel) and zero mean. These functions are described by adjustable hyper-parameters that define properties such as the variance, coherence scale, and shape. These hyper-parameters are optimised by maximising their posterior probability based on the observed data. Once the optimal model is obtained, the expected values of the foreground components are subtracted from the data. 

Following the standard GPR framework described by \citet{mertens_etal:2018}, the observed data $\mathbf{d}$ at frequencies $\nu$ are modelled by foreground $\mathbf{f}_\text{fg}$, 21-cm signal $\mathbf{f}_{21}$, and noise $\mathbf{n}$ components:
\begin{equation}
    \mathbf{d}(\nu) = \mathbf{f}_\text{fg}(\nu) + \mathbf{f}_{21}(\nu) + \mathbf{n}(\nu)\,.
\end{equation}
In our case, the $\mathbf{d}$ is the gridded $(u,\varv,\nu)$ visibility cube before making the power spectrum. This approach in the $u\varv$-space allows GPR to account for the baseline dependence of the frequency coherence scales, effectively modelling both the foreground wedge and thermal noise. By exploiting the distinct spectral behaviours of the different components, we can model the total covariance matrix as a sum of the individual GP covariance matrices:
\begin{equation}
    \mathbf{K}(\nu, \nu) = \mathbf{K}_\text{fg}(\nu, \nu) + \mathbf{K}_{21}(\nu, \nu) + \mathbf{K}_\text{n}(\nu, \nu)\, ,
\end{equation}
where $\mathbf{K}_\text{fg}$ represents the smooth foregrounds, $\mathbf{K}_{21}$ the 21-cm signal, and $\mathbf{K}_\text{n}$ the noise. The element $(p,q)$ of the covariance matrix corresponds to $\kappa(\nu_p, \nu_q)$, which is defined between two points $\nu_p$ and $\nu_q$, which, in our case, are two frequency channels (i.e.\ SB). Because $\textbf{n}$ is a Gaussian-distributed, frequency-uncorrelated noise with variance $\sigma^2_\text{n}$, the noise covariance matrix is $\textbf{K}_\text{n}(\nu, \nu) = \text{diag}[\sigma_\text{n}^2(\nu)]$. 

The foreground covariance includes two components: an intrinsic foreground $\mathbf{K}_\text{int}$ to capture the large frequency coherence scale of extragalactic and Galactic emissions, and a mode-mixing $\mathbf{K}_\text{mix}$ for the smaller frequency coherence scale ($1{-}5$\,MHz) of the foreground wedge. While the instrument chromaticity causes the mode-mixing to be a multiplicative effect, it can still be approximated as additive to first order, allowing us to define $\textbf{K}_\text{fg} = \textbf{K}_\text{int} + \textbf{K}_\text{mix}$, similar to \citet{mertens_etal:2020}.

As we will show in Section~\ref{sec:gpr:model}, the data can not be fully described by just the foreground and the 21-cm signal components. Similar to the NCP analysis \citep{mertens_etal:2020,munshi_etal:2024:ul}, also for the 3C\,196 data we observed an additional source of power with a small frequency coherence scale that is difficult to distinguish from the 21-cm signal. This `excess power' may arise from various instrumental effects, low-level RFI, polarization leakage, or calibration errors. We had to introduce an additional covariance matrix $\mathbf{K}_\text{ex}$ to capture the additional complexity of the data.

We can then define our total Gaussian process as the joint probability density distribution of the data $\textbf{d}$ and function values $\textbf{f}_\text{fg}$ of the foreground model at frequencies $\nu$:
\begin{equation}
    \begin{bmatrix}
        \mathbf{d} \\
        \mathbf{f}_\text{fg}
    \end{bmatrix}
    \sim \mathcal{N} \left( \,
    \begin{bmatrix}
        0 \\
        0
    \end{bmatrix}, 
    \begin{bmatrix}
        \mathbf{K}_{\text{fg}} + \mathbf{K}_{21} + \mathbf{K}_\text{ex} + \mathbf{K}_{\text{n}} & \mathbf{K}_{\text{fg}} \\
        \mathbf{K}_{\text{fg}} & \mathbf{K}_{\text{fg}}
    \end{bmatrix} \,\right)\, ,
\end{equation}
where we used the short-hand $\textbf{K} \equiv \textbf{K}(\nu, \nu)$. Because our data do not currently have the sensitivity for a detection of the 21-cm signal and our knowledge of the excess power is limited, we kept a conservative approach by subtracting only the modelled foreground components from the observed data:
\begin{equation}\label{eq:res_gpr}
    \textbf{d}_\text{res} = \textbf{d} - \mathbb{E}[\textbf{f}_\text{fg}]\, ,
\end{equation}
where $\mathbb{E}[\textbf{f}_\text{fg}]$ is the expectation value of the foreground model, given by
\begin{equation}\label{eq:gpr:expected_val}
    \mathbb{E}[\textbf{f}_\text{fg}]=\textbf{K}_\text{fg} [\mathbf{K}_{\text{fg}} + \mathbf{K}_{21} + \mathbf{K}_\text{ex} + \mathbf{K}_{\text{n}}]^{-1}\textbf{d} \, .
\end{equation}
The covariance of the foreground model is defined as
\begin{equation}\label{eq:gpr:covariance}
    \text{cov}[\textbf{f}_\text{fg}]= \textbf{K}_\text{fg} - \textbf{K}_\text{fg}[\mathbf{K}_{\text{fg}} + \mathbf{K}_{21} + \mathbf{K}_\text{ex} + \mathbf{K}_{\text{n}}]^{-1}\textbf{K}_\text{fg} \, .
\end{equation}
 
Equations~\eqref{eq:gpr:expected_val} and \eqref{eq:gpr:covariance} can be generalised to any component $\textbf{f}$ of our model to estimate its power spectrum and uncertainty. To achieve this, we draw $m$ samples from the posterior distribution of the hyper-parameters. For each sample, we compute $\mathbb{E}[\textbf{f}]$ and $\text{cov}[\textbf{f}]$ using the aforementioned equations. A power spectrum is then calculated by adding to $\mathbb{E}[\textbf{f}]$ a sample drawn from a Gaussian distribution with covariance $\text{cov}[\textbf{f}]$. The final power spectrum and its $1\sigma$ uncertainty are determined as the median and standard deviation of the $m$ power spectra at each $k$-mode.

\subsection{Covariance model}\label{sec:gpr:model}

Because the GPR model has zero mean, the components are completely defined by their covariance functions. Describing their correct form and shape is then fundamental. To model the foregrounds and the excess power, we found that the class of Matern covariance functions describes both smooth and rough frequency variations well. Its analytical form is given by \citep{stein:1999}
\begin{equation}\label{eq:matern_cov}
    \kappa_\text{Matern}(\nu_p, \nu_q) = \sigma^2\, \frac{2^{1-\mu}}{\Gamma(\mu)} \left( \frac{\sqrt{2\mu} \, r}{l} \right)^\mu K_\mu \left( \frac{\sqrt{2\mu} \, r}{l} \right)\, ,
\end{equation}
where $\sigma^2$ is the variance, $r = |\nu_q -\nu_p|$ is the absolute difference between the two frequencies, $l$ is the frequency coherence scale, $\mu$ is the smoothness parameter, $\Gamma$ is the Gamma function, and $K_\mu$ is the modified Bessel function of the second kind. Throughout this paper, all the $\sigma^2$ values are expressed as a fraction of the variance of the observed data. We used different kernels for the 21-cm signal and the thermal noise. Below, we describe the different covariance kernels used in our GPR model:

\paragraph*{{Intrinsic foreground -- \normalfont{$\mathbf{K}_\text{int}(\sigma^2_\text{int})$}:}} The intrinsic foreground is constituted by the residual extragalactic and Galactic emission within the field of view. Because it is expected to be spectrally smooth, we modelled the covariance function with a radial basis function (RBF), which is a Gaussian covariance function obtained by setting $\mu=\infty$ \citep{mertens_etal:2018}. It is the Matern function with the quickest drop at high $k_\|$, so it is ideal to model the residual foregrounds within the primary beam that are confined at low $k_\|$, as shown in Fig.~\ref{fig:PS2D_before_after_flag}. The function is characterised by two main hyper-parameters, the frequency coherence scale $l_\text{int}$ and the variance $\sigma_\text{int}^2$. Initially, we set a uniform prior $\mathcal{U}(10, 100)$\,MHz on $l_\text{int}$, but we found it was consistently unconstrained with a lower limit of ${\approx}60\,\text{MHz}$ after the optimisation. This indicated that $l_\text{int}$ was significantly larger than the data bandwidth (i.e.\ 12\,MHz), and the model was unable to recover the effective coherence scale. We therefore fixed $l_\text{int}$ to 80\,MHz, to reduce the number of degrees of freedom and speed up the fitting process. This choice did not affect the estimated values of the other hyper-parameters. 

\paragraph*{{Mode-mixing foreground} -- \normalfont{$\mathbf{K}_\text{mix}(\theta_\text{mix}, \sigma^2_\text{mix})$}:} Mode-mixing, caused by the chromatic response of the instrument, introduces smaller frequency coherence scales, typically in the range of 1--5 MHz. We found that using a Matern covariance function with $\mu=3/2$ provided the largest marginal likelihood (i.e.\ evidence) for the model. This is the same shape used by \citet{mertens_etal:2020}, where the two hyper-parameters $l_\text{mix}$ and $\sigma^2_\text{mix}$ were optimised. The result was a kernel with no dependence on baseline length, whereas mode-mixing effects should produce a wedge-like structure in the cylindrical power spectrum that depends on baseline length \citep{datta_etal:2010,morales_etal:2012}. To take into account this $k_\perp$ dependence of the scale length, we modified the Matern covariance function to parametrize such a wedge feature into the frequency coherence scale:
\begin{equation}\label{eq:gpr_wedgep}
    l_\text{mix}(u) = \frac{\nu_z}{\nu_z\,\eta_\text{buffer,mix} + u\sin\theta_\text{mix}}\, ,
\end{equation}
where $\eta_\text{buffer,mix}\propto k_\|$ is the delay buffer, $u\propto k_\perp$ is the baseline length (see Equation~\ref{eq:kperp_kpar}), $\theta_\text{mix}$ is the angle of the wedge-like structure in radians, and $\nu_z$ is the mean frequency of the redshift bin $z$. The delay buffer is used to ensure that the wedge does not start from $k_\| = 0$, adding a sort of extra intrinsic foreground. We found that $\eta_\text{buffer,mix}$ converged to $0.1\,\text{\textmu s}$ for our mode-mixing component. To decrease the number of degrees of freedom, we fixed this hyper-parameter to that value, leaving only $\theta_\text{mix}$ and $\sigma_\text{mix}$ to be optimised. We set a uniform prior $\mathcal{U}(0.01,1.6)$\,rad on $\theta_\text{mix}$, where the upper bound is ${\approx}\pi/2$ that is the maximum angle allowed for the horizon line. An issue with such a wedge parametrization is that Equation~\eqref{eq:gpr_wedgep} assumes the flat-sky horizon lines, but \citet{munshi_etal:2025} showed that the angle can be very different for phasing arrays such as LOFAR. An implementation of this improved wedge parametrization is left for the future. 

\paragraph*{{Excess power} -- \normalfont{$\mathbf{K}_\text{ex}(l_\text{ex}, \alpha_\text{ex}, \sigma^2_\text{ex})$}:} Most of our GPR efforts were put on finding the best kernel to describe the excess power. Even though recent works pointed out that such an excess is in large part related to DD gain errors on bright distant sources \citep{gan_etal:2022,brackenhoff_etal:2024,brackenhoff_etal:2025,ceccotti_etal:2025}, we do not know for certain its cause. Therefore, no prior knowledge was used to set up the excess power kernel shape. We found that the excess in our 3C\,196 data dominates the lower $k$-modes, in a `mode-mixing foreground'-like behaviour. While keeping all the other components unchanged and the same covariance function shape (we started with $\mu=\infty$, i.e.\ an RBF kernel), we saw that by using the wedge parametrization the marginal likelihood of the GPR model increased by a few per cent. However, for the excess, $\theta$ was giving $\pi/2$ as a lower limit. Because that is the maximum angle allowed by Equation~\eqref{eq:gpr_wedgep}, and we know that the real horizon limit has an angle larger than the flat-sky horizon line, we had to test a different parametrization to push this limit. Instead of using an angle $\theta$, we can express the $k_\perp$ dependence of the coherence scale and the variance as
\begin{equation}\label{eq:gpr_alpha}
    l_\text{ex}(u) = \frac{l_\text{0,ex}}{1+10^{-3}\alpha_\text{ex} (u-u_\text{min})}\, ,\ \sigma_\text{ex}^2(u) = \sigma_{0,\text{ex}}^2 \left(\frac{u}{u_\text{min}}\right)^{\alpha_{\sigma,\text{ex}}}\, ,
\end{equation}
where $l_\text{0,ex}$ is the coherence scale at $u=u_\text{min}$, $u_\text{min}=50\lambda$ is the minimum baseline length, and $\alpha_\text{ex}$ and $\alpha_{\sigma,\text{ex}}$ are the coefficients that encode the baseline dependence into the coherence scale and the variance, respectively. Given the lack of prior knowledge on the excess kernel, we set a uniform prior $\mathcal{U}(-5,60)$ on $\alpha_\text{ex}$, where the large upper bound was to allow a steep wedge (even above the expected real horizon limit), while the lower negative bound was to rule out an inverse wedge-like structure. For $l_\text{0,ex}$, we set a uniform prior $\mathcal{U}(0.2, 0.8)\,\text{MHz}$, which gives a very small coherence scale. Keeping the same prior ranges, we found that using a Matern covariance function with $\mu =5/2$ gave higher marginal likelihood than an RBF and $\mu=3/2$. While we would expect a flat baseline dependence for the variance, the data preferred $\alpha_{\sigma,\text{ex}}<0$. To decrease the degrees of freedom, we fixed this parameter to its converging value, namely $\alpha_{\sigma,\text{ex}}=-0.25$. We report some of the most relevant tests in Appendix~\ref{app:excess_tests}. 

\paragraph*{{21-cm signal} -- \normalfont{$\mathbf{K}_\text{21}(x_1, x_2, \sigma^2_{21})$}.} Instead of using a Matern covariance function, the 21-cm signal kernel was constructed using an ML-based variational auto-encoder (VAE) kernel, as described by \citet{mertens_etal:2024} and \citet{acharya_etal:2024}. The VAE kernel is trained on simulations to compress the high-dimensional 21-cm signal data into a lower-dimensional latent space, from which the covariance matrix is reconstructed. This approach was necessary because standard GPR kernels could struggle to capture the complex frequency covariance of the 21-cm signal in the observed data, leading to biases and risk of signal loss in the power spectrum estimation \citep{kern_liu:2021}. The VAE kernel, in contrast, adapts to the different physical characteristics of the signal learned from 21-cm signal simulations. This ML-GPR approach has already been applied by \citet{munshi_etal:2024:ul}. 

For this work, the VAE kernel was trained on 21-cm signal models for $z=9.16$, using the simulation framework described by \citet{acharya_etal:2024}. The EoR simulations were generated with \textsc{grizzly} \citep{ghara_etal:2015a,ghara_etal:2015b,ghara_etal:2018,ghara_etal:2020}, a 1D radiative transfer code coupled with cosmological and $N$-body simulations to produce 21-cm brightness temperature maps at different redshifts. The ML-trained kernel resulted in a covariance function described by only two latent space dimensions, $x_1$ and $x_2$, and a scaling factor for the 21-cm signal variance $\sigma_{21}^2$. During the training, the VAE forces the latent space to be normally distributed with variance one. We set a uniform prior $\mathcal{U}(-3,3)$ on $x_1$ and $x_2$, a choice that is sufficient to explore any possible expected shape of the 21-cm signal power spectra.

\paragraph*{{Thermal noise} -- \normalfont{$\mathbf{K}_\text{n}$}:}  The noise covariance function is built from the time-differenced Stokes~V visibilities, which can be slightly lower than the noise in Stokes~I. Therefore, we estimated the scaling factor using an initial uniform prior $\mathcal{U}(0.5, 2.5)$ and consistently found a value of one across different tests. We then fixed the scaling factor to this value to reduce the number of degrees of freedom and reduce the computing time.


\subsection{Application to data and residual power spectrum}\label{sec:gpr:application}

\begin{table*}
\centering
\caption{Components of our ML-GPR model, along with the parameter priors and the estimated median values with the $1\sigma$ uncertainties. All the variances are expressed as a fraction of the input data.}
\label{table:param_estimation}
\resizebox{1\textwidth}{!}{
\begin{tabular}{llclcc}
\toprule
Component & Covariance & Parameter & Description & Prior & Estimated value \\ \midrule\midrule
Intrinsic foregrounds & Radial Basis Function &  $\sigma^2_{\text{int}}$ & Variance & $\log\mathcal{U}(-0.5, 0.5)$ & $-0.38 \pm 0.02$ \\ \midrule
Mode-mixing foregrounds & Matern Function ($\eta = 3/2$) & $\sigma^2_{\text{mix}}$ & Variance & $\log\mathcal{U}(-0.8, -0.01)$ & $-0.34 \pm 0.01$ \\
& & $\theta_{\text{mix}}$ & Angle (rad) & $\mathcal{U}(0.01, 1.6)$ & $0.175 \pm 0.003$ \\ \midrule
21-cm signal & Trained VAE-based Kernel & $x_1$ & Latent space dimension & $\mathcal{U}(-3, 3)$ & -- \\
& & $x_2$ & Latent space dimension & $\mathcal{U}(-3, 3)$ & -- \\
& & $\sigma^2_{21}$ & Variance & $\log\mathcal{U}(-7, -0.5)$ & $< -3.2$ \\ \midrule
Excess power & Matern Function ($\eta = 5/2$) & $\sigma^2_{0,\text{ex}}$ & Variance & $\log\mathcal{U}(-4, -2)$ & $-2.51 \pm 0.2$ \\
& & $l_{0,\text{ex}}$ & Length scale (MHz) & $\mathcal{U}(0.2, 0.8)$ & $0.52 \pm 0.02$ \\
& & $\alpha_\text{ex}$ & Baseline dependence & $\mathcal{U}(-5, 60)$ & $36 \pm 1$ \\
\bottomrule
\end{tabular}}
\end{table*}

The input data for our ML-GPR model consists of the gridded visibility cubes after the outlier flagging, as described in Section~\ref{sec:proc:imaging}. The posterior probability distributions and the Bayesian evidence of the covariance model were derived with the \textsc{ultranest}\footnote{\url{https://johannesbuchner.github.io/UltraNest}} package \citep{buchner_etal:2021}, based on a nested sampling Monte Carlo algorithm \citep{buchner_etal:2016,buchner_etal:2019}. We used 100 live points to explore the parameter space within the prior constraints and find the posterior distribution of the hyper-parameters. 

\begin{figure}
    \centering
    \includegraphics[width=1\columnwidth]{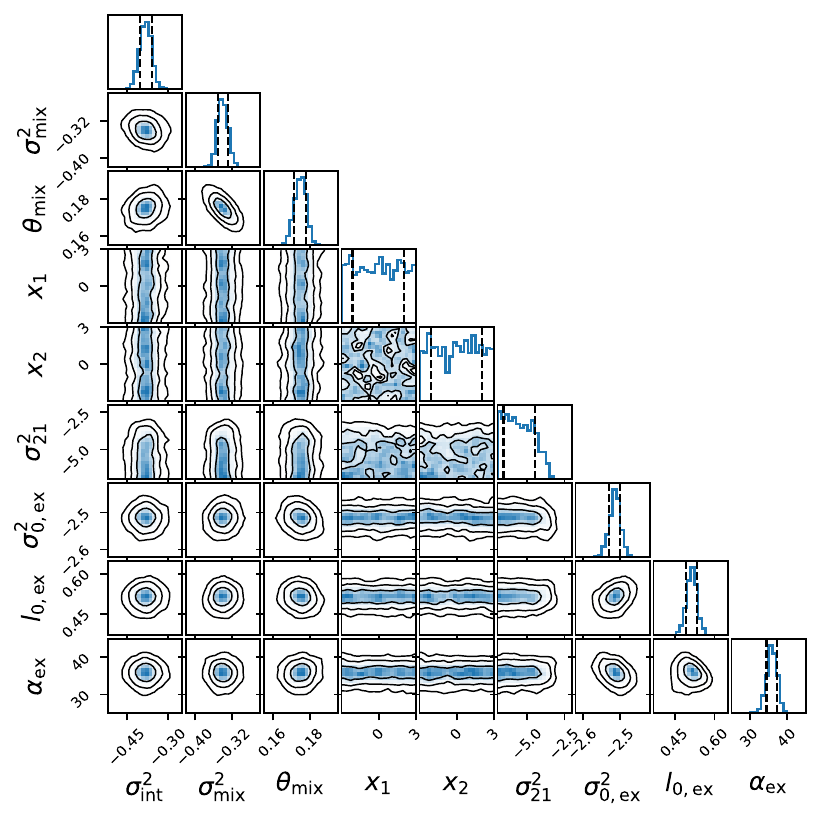}
    \caption{Posterior probability distribution for each of the ML-GPR hyper-parameters (Table~\ref{table:param_estimation}). The contours represent the 68, 95, and 99.7 per cent confidence intervals. The vertical dashed lines in the histograms indicate the $16^\text{th}$ and the $84^\text{th}$ percentile of the distributions.}
    \label{fig:ML-GPR_corner}
\end{figure}

The prior ranges and the results of the parameter estimation are presented in Table~\ref{table:param_estimation}, with a corner plot of the posterior distributions shown in Fig.~\ref{fig:ML-GPR_corner}. The posterior distributions of the foregrounds and excess parameters are peaked and symmetric around the estimated values, giving small uncertainties. Parameters of the same component show low to moderate correlation, with a high negative correlation between $\sigma_\text{mix}^2$ and $\theta_\text{mix}$. Because the 21-cm signal component is well below the thermal noise level, its parameters $x_1$ and $x_2$ did not converge and give all the allowed signal shapes equally probable, as should be expected. For the same reason, its variance $\sigma_{21}^2$ reached an upper limit.

\begin{figure}
    \centering
    \includegraphics[width=1\columnwidth]{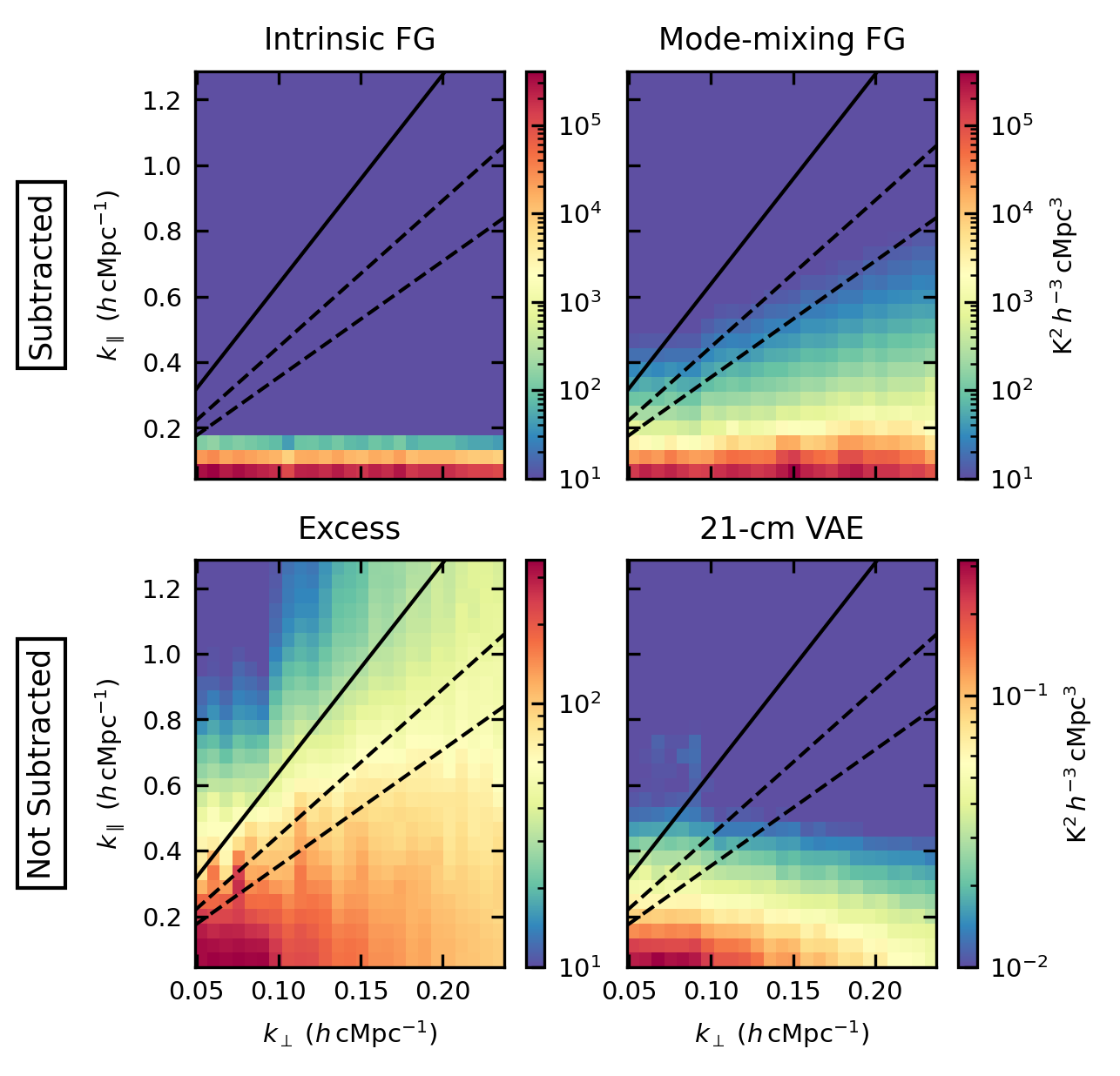}
    \caption{Cylindrical power spectra of the components of our ML-GPR model: intrinsic foreground (top left), mode-mixing component (top right), excess power (bottom left), and 21-cm signal (bottom right). The top row shows the components that are subtracted from the observed data, whereas the bottom row shows the components that will form the residual data along with the thermal noise.  In all panels, the solid line indicates the horizon delay line, whereas the dashed lines indicate the delay ranges where we expect most of the power of Cas\,A. The same colour range is used for the intrinsic and the mode-mixing foregrounds.}
    \label{fig:ML-GPR_kernels_best}
\end{figure}

From these posterior distributions, we sampled 500 realisations of each component cube. For each realisation, we estimated a power spectrum and we took the median of these 500 spectra at each $k$-mode. The resulting cylindrical power spectra $P(k_\perp, k_\|)$ are shown in Fig.~\ref{fig:ML-GPR_kernels_best}. While intrinsic and mode-mixing foregrounds are similar to the NCP results, the excess component for our 3C\,196 data shows a foreground-like feature, in contrast to a more noise-like behaviour for the NCP excess \citep{mertens_etal:2020}. Our excess is characterised by a wedge structure confined within the real horizon limit (solid black line), given by $\alpha_\text{ex}=36\pm 1$. The peak in power observed in the bottom-left corner of the $k_\perp\text{-}k_\|$ space suggests that the source of such an excess may be related to residual extended emission outside the primary beam. A few bright ($k_\perp, k_\|$)-cells are located along the Cas\,A direction (dashed black lines), where the DD-subtracted data showed some residual emission (see Fig.~\ref{fig:UVplane_DI_DD}). The kernel large extent in $k_\|$ causes some leakage into the EoR window, where the power is approximately an order of magnitude higher than the modelled mode-mixing foreground. However, as we will discuss later, the contamination in the EoR window remains at or below the thermal noise level. The residual data cube was then obtained by subtracting the sampled realisations of the two foreground components from the input data (Equation~\ref{eq:res_gpr}). 

\begin{figure}
    \centering
    \includegraphics[width=1\columnwidth]{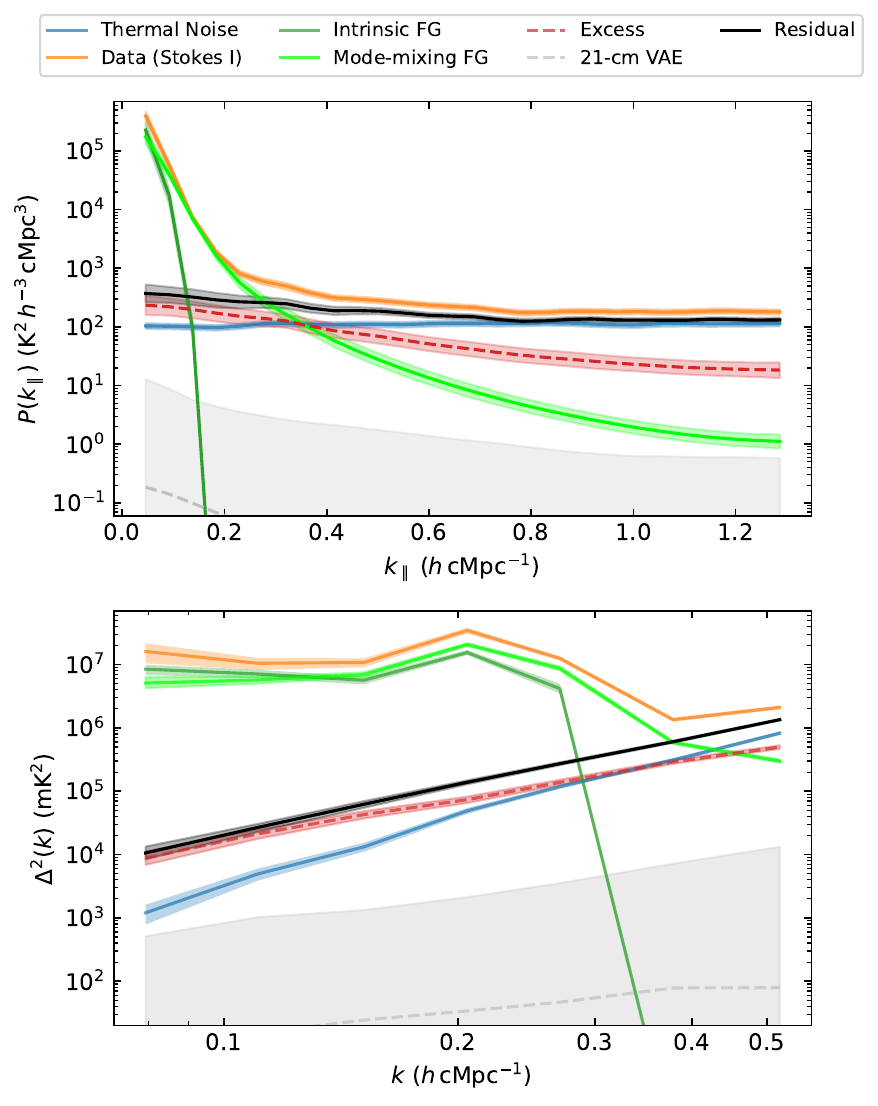}
    \caption{Decomposition of the components of our ML-GPR model for the input residual Stokes~I data (orange solid line). The covariance model is constituted by intrinsic foreground (green solid line), mode-mixing component (lime solid line), excess (red dashed line), 21-cm signal (gray dashed line), and thermal noise (blue line). The ML-GPR residual data (black solid line) are obtained by subtracting the foreground components from the input data. The $2\sigma$ uncertainties are reported with the shaded areas. The top panel shows the cylindrical power spectra as a function of $k_\|$ (i.e.\ averaged over $k_\perp$), whereas the bottom panel shows the spherical power spectra.}
    \label{fig:PS1D_GPR_results}
\end{figure}

A summary of the power spectra of input data, ML-GPR components, and residual data is shown in Fig.~\ref{fig:PS1D_GPR_results}, where the $k_\perp$-averaged power spectrum $P(k_\|)$ and the spherical power spectrum $\Delta^2(k)$ are shown in the top and bottom panel, respectively. The excess component is higher than the noise at $k\approx k_\|\lesssim 0.3\,h\,\text{cMpc}^{-1}$, but goes below it at higher $k$, becoming a fraction of the noise power. This behaviour resembles the power spectrum of a foreground component but with much smaller spectral coherence. In these plots we also show the 21-cm signal component, whose variance was at least more than two orders of magnitude lower than the data variance and the parameters $x_1$ and $x_2$ were unconstrained. This results in a genuine upper limit on the 21-cm signal power spectrum with albeit large uncertainties, consistent with \citet{mertens_etal:2024}. 

\begin{figure}
    \centering
    \includegraphics[width=0.92\columnwidth]{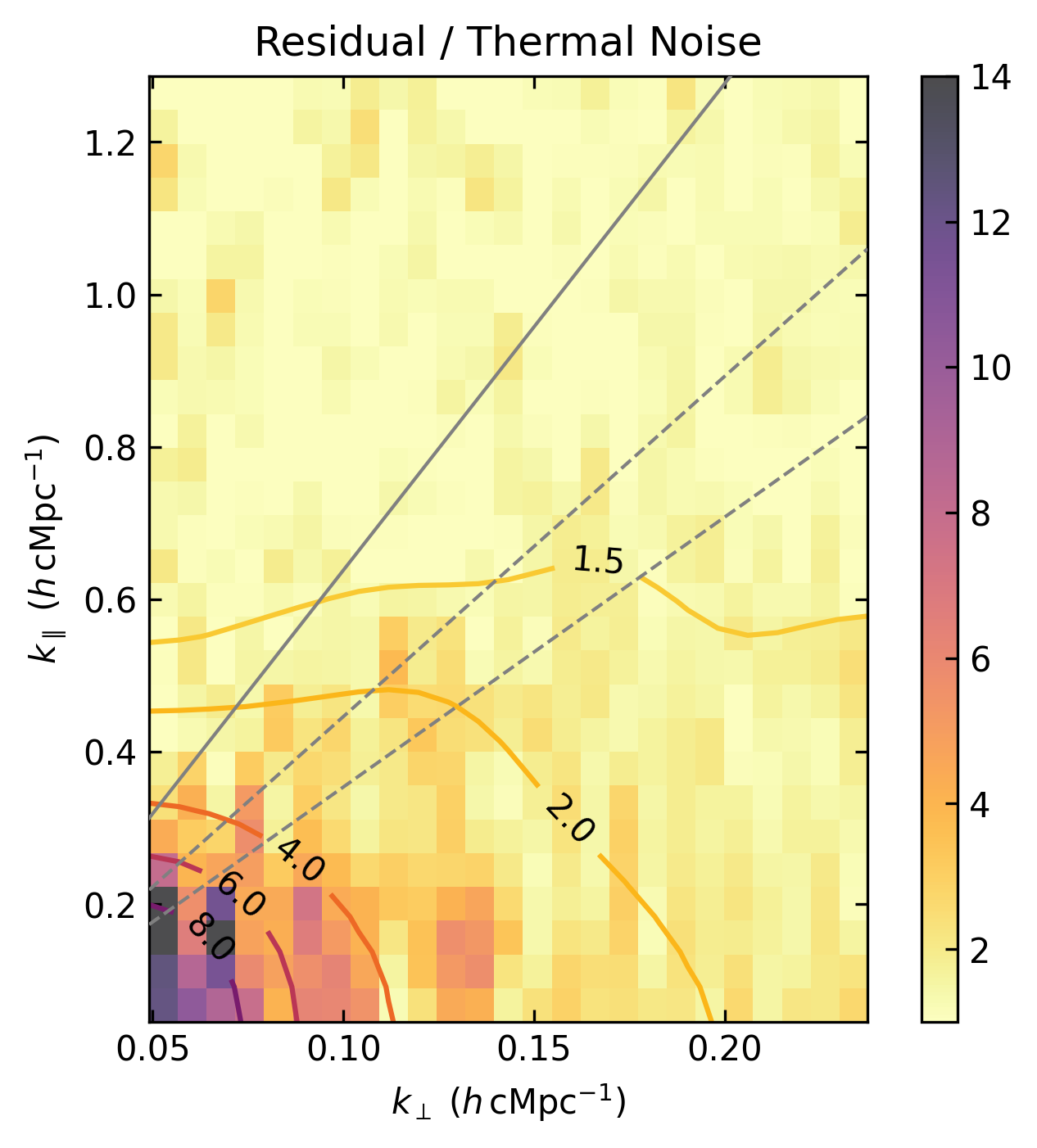}
    \caption{Cylindrical power spectrum of the ML-GPR residuals divided by the power spectrum of the thermal noise (i.e.\ time-differenced Stokes~V). The grey solid line indicates the horizon delay line, whereas the grey dashed lines indicate the delay ranges where we expect most of the power of Cas\,A. The colour range between 1 and 14 is chosen to allow an easier comparison with Fig.~12 in \citet{mertens_etal:2020} .}
    \label{fig:PS2D_ratio_rescNoise}
\end{figure}

Figure~\ref{fig:PS1D_GPR_results} also shows the residual power spectra. Because only the foreground components were subtracted from the input data, the residual power spectrum mainly consists of the excess at low $k_\|$ and noise at high $k_\|$. The ratio between the cylindrical power spectra of residual and thermal noise is shown in Fig.~\ref{fig:PS2D_ratio_rescNoise}. The EoR window is very clean, with a mean ratio of approximately 1.2. On the other hand, the ratio is approximately 3.0 on average within the wedge, being dominated by the excess component especially at $k_\perp < 0.11\,h\,\text{cMpc}^{-1}$, where the mean ratio is approximately $5.8$. Note that the noise power spectrum estimated from time-differenced Stokes~V is just a realisation drawn from the theoretical noise distribution, which is expected to have a constant power at fixed $k_\perp$. This means that every $k_\|$-cell at fixed $k_\perp$ should have a value more or less close to the mean. However, there are a few ($k_\perp$, $k_\|$)-cells where the observed thermal noise is more than $2\sigma$ lower or higher than the mean value, resulting in a quite high or low ratio, respectively. This is the case for the very high ratios at $k_\perp\lesssim0.07\,h\,\text{cMpc}^{-1}$ and $k_\| \approx 0.15\,h\,\text{cMpc}^{-1}$.

\section{Validation of the processing pipeline}\label{sec:validation}

Because the foreground emission is several orders of magnitude brighter than the 21-cm signal, the data calibration and foreground removal steps might suppress or bias the 21-cm results. While the power spectrum estimation pipeline has been extensively tested against known power spectra \citep{mertens_etal:2020}, the calibration steps and the ML-GPR foreground removal need to be validated against signal loss. 

\subsection{Robustness test on calibration steps} 

The main danger of the DI-calibration (Section~\ref{sec:proc:di-cal}) is setting an incorrect flux scale, biasing the final results by a scaling factor. The bottom panel of Fig.~\ref{fig:3C196_model_flux_afterDI} shows that the peak brightness of 3C\,196 after the DI-steps (black dots) agrees with the expected total flux of the source \citep{scaife_heald:2012}. This ensures that the flux scale is correctly set for the DI-calibrated data. The DD solving and sky model subtraction (Section~\ref{sec:proc:dd-cal}) could affect the resulting power spectra more drastically because they have the potential to modify and remove part of the 21-cm signal \citep{patil_etal:2016,ewall-wice_etal:2017}. This bias is strongly curtailed by solving gains using only baselines longer than $250\lambda$, as shown by \citet{mevius_etal:2022}. Moreover, we applied a smoothing kernel of 4\,MHz width, such that over-fitting and 21-cm signal suppression are reduced. This smoothing is different from the standard method employed in the NCP processing \citep{patil_etal:2017,mertens_etal:2020,mertens_etal:2025}, which makes use of \textsc{sagecal} \citep{yatawatta:2016}. Instead of smoothing the solution with a Gaussian kernel at each iteration, as done in \textsc{ddecal}, \textsc{sagecal} uses a consensus optimisation algorithm to spectrally constrain the solutions with a third-order Bernstein polynomial \citep[more details in][]{yatawatta:2015,yatawatta:2016,yatawatta:2017,yatawatta:2018}. \citet{gan_etal:2023} showed that \textsc{ddecal} with our settings gives results on the 21-cm power spectrum comparable to \textsc{sagecal}, implying limited signal loss in the DD calibration step. 

\subsection{Robustness test on ML-GPR} 

The GPR signal-separation method has been extensively tested, and its reliability in statistically separating foreground and 21-cm signal was confirmed \citep{mertens_etal:2018,offringa_etal:2019a}. However, any small error in the foreground components might completely alter the inferred 21-cm signal because of their up to five orders of magnitude difference in power. The new ML-GPR method limits this problem by using an ML-based VAE kernel, but a bias in the excess component might still happen. 

The validation process of the ML-GPR foreground removal is similar to that by \citet{mertens_etal:2020}, where synthetic 21-cm signals are generated and added to the data. We performed such a signal injection test with the following steps:
\begin{enumerate}
    \item \textit{Signal generation}: The trained 21-cm VAE kernel was used to generate 25 different power spectrum shapes by uniformly sampling $x_1$ and $x_2$ between $-2$ and $2$. Each power spectrum shape was then scaled to a power equal to $0.5$, $1$, and $2$ times the thermal noise power $\sigma_\text{n}^2$, resulting in 75 unique mock 21-cm signals. A Gaussian random realisation of a 21-cm signal gridded visibility cube was generated per signal shape and intensity.

    \item \textit{Injection}: The simulated 21-cm visibility cube was added to the gridded data cube after the outlier flag, just before performing the ML-GPR step.

    \item \textit{ML-GPR application}: ML-GPR was subsequently applied to the data cube with the injected signal using the same priors as for the original observed data. A new set of optimal foreground, excess and 21-cm kernels was found for each injected signal.

    \item \textit{Recovered signal estimation}: Residual power spectra were calculated as in Equation~\eqref{eq:res_gpr}. A recovered 21-cm power spectrum $\Delta_\text{rec}^2$ was then estimated by subtracting the residual power spectrum of the original data (without the injected signal) from the residual power spectrum of the data with the injected signal.

    \item \textit{Comparison}: Each recovered 21-cm power spectrum was compared to the power spectrum of the corresponding injected signal $\Delta_\text{inj}^2$ using the two statistical estimators:
    \begin{itemize}
        \item $z$-score: It describes the deviation of the recovered signal from the injected signal in units of standard deviations at each $k$-bin. The $z$-score was estimated as the inverse cumulative distribution function of the normal distribution, given a $p$-value for each $k$-bin calculated as the fraction of realisations where $\Delta_\text{rec}^2(k) > \Delta_\text{inj}^2(k)$ (our null hypothesis). With this null hypothesis, $z\text{-score}<0$ indicates signal absorption, with $z\text{-score}<-2$ suggesting a suppression beyond the $2\sigma$ upper limit.\\

        \item Bias: It measures the factor by which the recovered signal differs from the injected signal at each $k$-bin, calculated as
        \begin{equation}
            b_k = \frac{\Delta_\text{rec}^2(k)}{\Delta_\text{inj}^2(k)}\,.
        \end{equation}
        A $z\text{-score}<0$ means $b_k<1$, and the employed ML-GPR model might bring signal loss. 
    \end{itemize}
\end{enumerate}

\begin{figure}
    \centering
    \includegraphics[width=1\columnwidth]{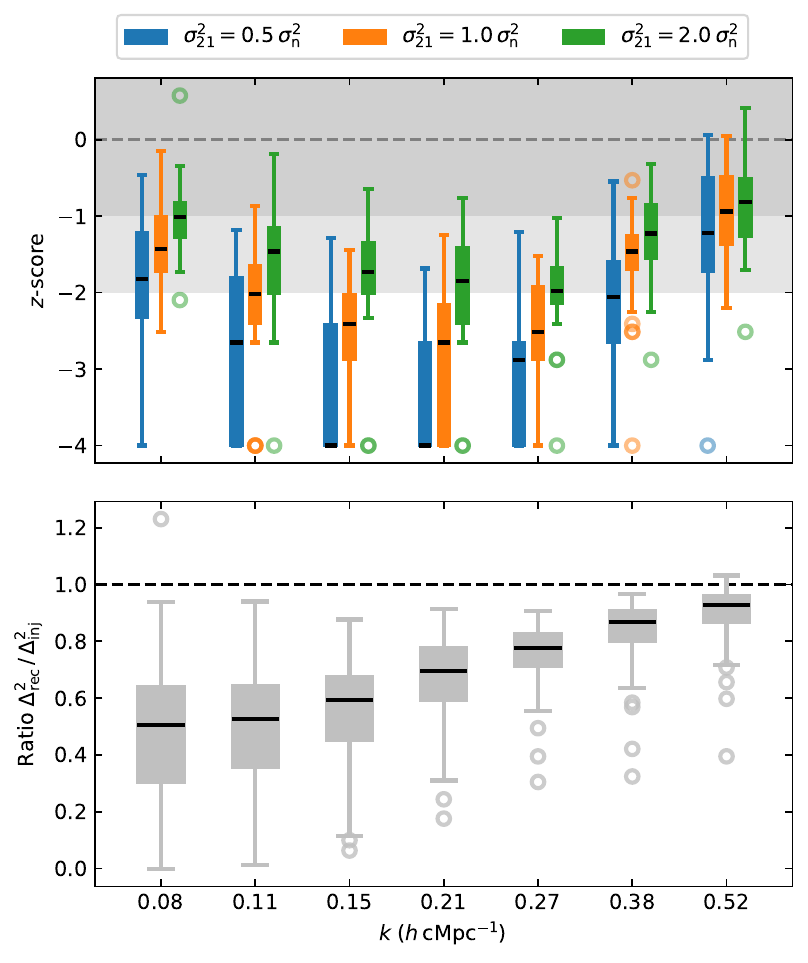}
    \caption{Box plots of the $z$-scores (top) and bias factor $b_k=\Delta_\text{rec}^2 / \Delta_\text{inj}^2$ (bottom) as a function of $k$-mode, resulting from the injection test. Each box represents the inter-quartile range (IQR), with the median indicated by a black solid line, the whiskers extending to $1.5\,\text{IQR}$ from the box, and outliers shown as circles. We injected 25 mock 21-cm signals of different shapes, and scaled their intensities to match 0.5 (blue), 1 (orange), and 2 (green) times the noise variance. In the top panel, each box plot shows the $z$-score distributions for all the 25 shapes, with different colours for the different intensities. The dark and light grey shaded areas indicate the $1\sigma$ and $2\sigma$ levels, respectively. In the bottom panel, each box plot represents the distribution of the bias factor for all the 75 injected signals.}
    \label{fig:ML-GPR_inj_zscore_ratio}
\end{figure}

\begin{figure*}
    \centering
    \includegraphics[width=1\textwidth]{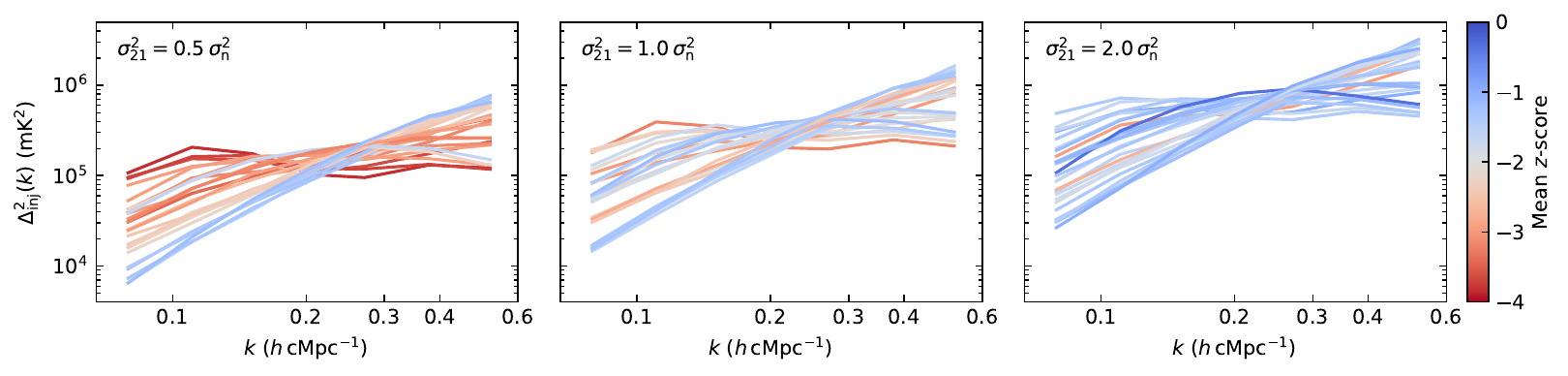}
    \caption{Spherical power spectra of the 25 injected 21-cm signal shapes. From left to right, the panels show the same signal shapes with different intensities, namely 0.5, 1, and $2\,\sigma_\text{n}^2$. Colours indicate the mean $z$-score, averaged over the $k$-modes.}
    \label{fig:ML-GPR_inj_pseor}
\end{figure*}

The $z$-scores for all the 75 signal injection tests are shown in the top panel of Fig.~\ref{fig:ML-GPR_inj_zscore_ratio}. Our ML-GPR model was not able to recover most of the injected signals, resulting in approximately 62, 47, and 24 per cent of $z\text{-scores}<-2$ for $\sigma_{21}^2/\sigma_\text{n}^2=0.5$, 1, and 2, respectively. Most of these are at $0.11\le k\le 0.27\,h\,\text{cMpc}^{-1}$, while at the edges of the samples $k$-space the median $z$-score for all the intensities is within the $2\sigma$ upper limit. We also observe a trend with the intensity, where higher $\sigma_{21}^2$ resulted in better recovered signals. This is shown in Fig.~\ref{fig:ML-GPR_inj_pseor} with the power spectra of the different injected signal shapes. The colours indicate the mean $z$-score over $k$ for each combination of $x_1$ and $x_2$. Most of the cases with $z\text{-scores}<-2$ show a flat power spectrum or an upturn at low $k$. These are more easily absorbed into the mode-mixing and excess components because of their larger spectral coherence scale. Steeper power spectra or with a bump around $0.2\,h\,\text{cMpc}^{-1}$ were better recovered, being less similar to a foreground component. The fact that injected signals with higher variance were better recovered might also mean that ML-GPR is not successful in modelling data with strong foreground residuals. Because of the limited spatial extent of our sky model and lack of extended emission, our ML-GPR input data have high power close to the primary beam null and in the sidelobes (see Fig.~\ref{fig:3C196_DI_DD_widefield_zoomin}, Fig.~\ref{fig:3C196_afterDD_StokesI_V}, and Fig.~\ref{fig:PS2D_before_after_flag}). This brought our ML-GPR model to be able to differentiate between the 21-cm signal and the mode-mixing component only when the injected signal is higher than the mode-mixing for most of the $k$-bins. 

The bottom panel of Fig.~\ref{fig:ML-GPR_inj_zscore_ratio} shows the bias factor $b_k$ for each $k$-bin. While a few tests returned $b_k \leq 1$ at $k = 0.08$ and $0.52\,h\,\text{cMpc}^{-1}$, the median value is below one for all $k$-bins, as expected from the $z$-scores. This bias indicates that our ML-GPR model would likely suppress the 21-cm signal, especially when it is at or below the noise level. Therefore, a bias $b_k < 1$ results in lower upper limits, potentially leading to incorrect interpretation. To correct for this bias, we can divide the final power spectrum results by $b_k$ at each $k$-bin, giving conservative upper limits.

\subsection{Data and model power spectra}\label{sec:validation:data-model}

\begin{figure*}
    \centering
    \includegraphics[width=1\textwidth]{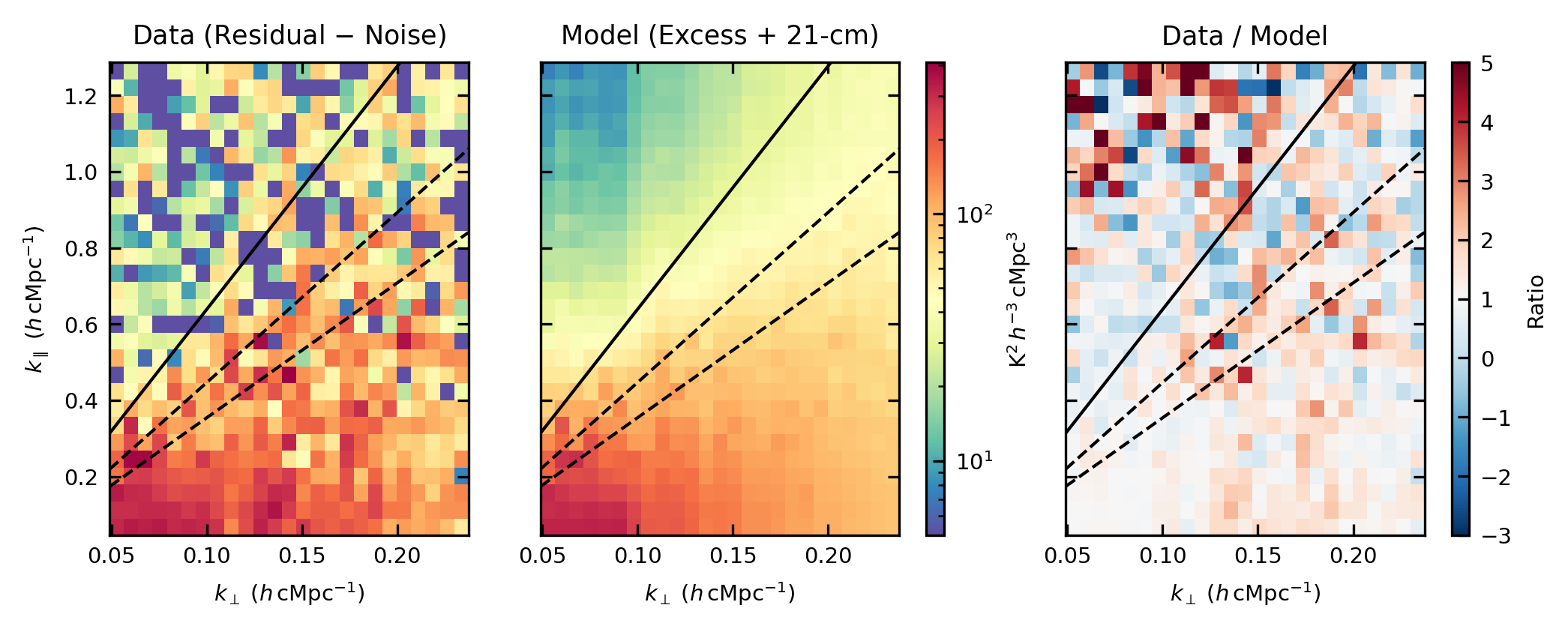}
    \caption{Cylindrical power spectra of the noise-subtracted data residuals (first panel), model residuals (second panel), and ratio between them (third panel). The same colour bar is used for the first and second panels. In all panels, the black solid line indicates the horizon delay line, while the black dashed lines indicate the delay ranges where we expect most of the power of Cas\,A.}
    \label{fig:PS2D_residual_data_model}
\end{figure*}

In addition to the strong foreground residuals, another possible reason for larger biases could be GPR model incompleteness. If there is an extra component not modelled by our set of kernels, adding a signal on top of it might lead to a mix of this component with the injected signal, preventing full recovery of the latter. To ensure this is not the case, we estimated the `model residual` as
\begin{equation}\label{eq:validation:model_res}
    \textbf{m}_\text{res} = \mathbb{E}[\textbf{f}_\text{ex}] + \mathbb{E}[\textbf{f}_{21}]\, ,
\end{equation}
and compared its power spectrum with the noise-subtracted power spectrum of the data residuals, given by
\begin{equation}\label{eq:validation:data_res}
    P_\text{d} = P_\text{I,res} - P_\text{n}.
\end{equation}
If the ML-GPR model is complete and our kernel set fully describes the data, these two estimators should match. The cylindrical power spectra of the data and model residuals are shown in the first and second panels of Fig.~\ref{fig:PS2D_residual_data_model}, respectively. The data power spectrum is obtained by subtracting the thermal noise power spectrum from the ML-GPR residual data. This operation results in some ($k_\perp$-$k_\|$)-cells with negative power, particularly in the EoR window, where the residual data variance is dominated by the noise variance (see Fig.~\ref{fig:PS2D_ratio_rescNoise}).

The ratio of the two power spectra is plotted in the third panel of Fig.~\ref{fig:PS2D_residual_data_model}. In the bottom-left corner, where the excess component dominates the residuals, the ratio is around 1. At higher $k_\perp$, within the wedge up to the Cas\,A direction, the data residuals show more power than the model residuals. Above the Cas\,A delay lines, there are regions where the model residuals are stronger, but these mainly correspond to the ($k_\perp$-$k_\|$)-cells where the noise variance dominates over the excess. The ratio suggests that a component with higher power than the excess at $k_\perp > 0.12\,h\,\text{cMpc}^{-1}$ and within the wedge might be necessary. We tested an excess kernel with $\alpha_{\sigma,\text{ex}} = 0$, and the resulting evidence was lower than for $\alpha_{\sigma,\text{ex}} = -0.25$, indicating that our excess kernel did not better fit the data with higher variance at longer baselines. However, \citet{mertens_etal:2025} showed that using two mode-mixing components with different $\theta_\text{mix}$ improved results at $z=10.1$, where the foreground is brighter. We do not exclude the possibility that adding an extra excess or mode-mixing component with higher variance at larger $k_\perp$ could reduce the difference between the data and model residuals and ultimately allow the injection tests to pass with $z{\text{-score}} > -2$. Nonetheless, since the two estimators agree within their uncertainties (see the spherical power spectra in Section~\ref{sec:results}), we conclude that our ML-GPR model is mostly complete in the $k$-bins of interest, with no significant indication of missing components.

\section{Results and upper limits}\label{sec:results}

We finally derived the upper limits on the 21-cm signal from the ML-GPR Stokes~I residual data. The spherical power spectrum is estimated in seven $k$-bins logarithmically spaced between $0.06$ and $0.5\,h\,\text{cMpc}^{-1}$, with a bin size of $\Delta k / k\approx 0.3$. Similar to \citet{mertens_etal:2020}, we subtracted the noise bias from the Stokes~I residual power spectrum $\Delta_{\text{I,res}}^2$. We used the time-differenced Stokes~V, the same as given as input in the ML-GPR model, to estimate the noise bias power spectrum $\Delta_\text{n}^2$. The spherical noise-subtracted power spectrum of the residual is then defined from Equation~\eqref{eq:validation:data_res} as
\begin{equation}\label{eq:21cm_data}
    \Delta_{21,\text{d}}^2 = \Delta_{\text{I,res}}^2 - \Delta_\text{n}^2\,.
\end{equation}
Similar to Section~\ref{sec:validation:data-model}, we compare $\Delta_{21,\text{d}}^2$ with the spherical power spectrum of the model residual $\Delta_{21,\text{m}}^2$, estimated from the visibility cube of Equation~\eqref{eq:validation:model_res}. The uncertainties of the power spectra are estimated by using a sampling approach within our ML-GPR framework. By generating multiple realizations of the power spectrum through sampling of the hyper-parameter posterior distributions, we obtained a distribution for each $k$-bin, as described in Section~\ref{sec:gpr:application}. We then computed the $97.5^\text{th}$ percentile $\mathcal{P}_{97.5}$ of these distributions, providing the $2\sigma$ upper limit at each $k$-bin. 

\begin{figure}
    \centering
    \includegraphics[width=1\columnwidth]{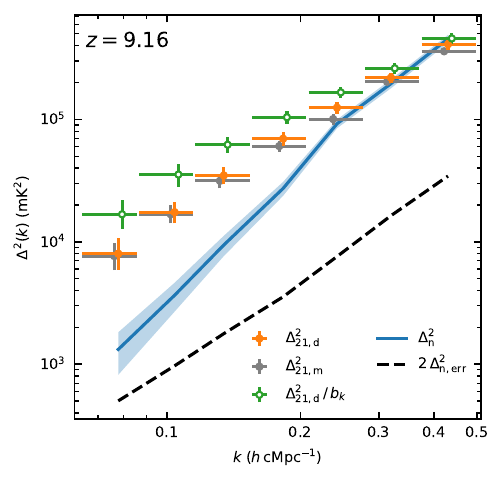}
    \caption{Final Stokes~I spherical power spectra after ML-GPR. We plot both the noise-subtracted power spectrum of the data residuals (orange points) and the power spectrum of the model residuals (grey points). The green circles represent the bias-corrected power spectrum, which is used to estimate the upper limits on the 21-cm signal. The thermal noise level is shown with the blue line. The horizontal error bars indicate the $k$-bin extension, while the vertical ones and the blue shaded area indicate the 95 per cent confidence interval (i.e.\ $2\sigma$ uncertainties). A small offset in $k$ has been added to the grey points and green circles to avoid overlapping of error bars. The black dashed line represents the lowest achievable upper limit (given the injection test bias) if the ML-GPR residuals were fully thermal noise dominated, namely the $2\sigma$ uncertainties of the thermal noise.}
    \label{fig:PS3D_data_model_biascorr}
\end{figure}

\begin{table*}
\centering
\caption{Spherical power spectrum values at each $k$-bin for the different 21-cm signal estimators, namely the noise-subtracted residual data  $\Delta_{21,\text{d}}^2$, the model residual $\Delta_{21,\text{m}}^2$, and the bias-corrected residual data $\Delta_{21,\text{d}}^2/b_k$, that we consider for our final results. The upper limits are reported as the $97.5^\text{th}$ percentile of the power spectrum distributions. The last two columns report the power spectrum of the thermal noise and its $2\sigma$ uncertainty.}
\label{table:delta_21_ul}
\begin{tabular}{lccccccccc}
\toprule
$k$ & $\Delta_{21,\text{d}}^2$ & $\mathcal{P}_{97.5}(\Delta_{21,\text{d}}^2)$ & $\Delta_{21,\text{m}}^{2}$ & $\mathcal{P}_{97.5}(\Delta_{21,\text{m}}^{2})$ & $\mathcal{P}_{97.5}(\Delta_{21,\text{d}}^{2}/b_k)$ & $\Delta_\text{n}^2$ & $2\Delta_{\text{n,err}}^2$ \\ 
$(h\,\text{cMpc}^{-1})$ & $(\text{mK}^2)$ & $(\text{mK}^2)$ & $(\text{mK}^2)$ & $(\text{mK}^2)$ & $(\text{mK}^2)$ & $(\text{mK}^2)$ & $(\text{mK}^2)$ \\
\midrule
$0.078$ & $(89.14)^2$ & $(100.99)^2$ & $(87.26)^2$ & $(98.72)^2$ & $(146.61)^2$ & $(36.33)^2$ & $(22.40)^2$ \\ 
$0.104$ & $(131.73)^2$ & $(145.16)^2$ & $(129.05)^2$ & $(141.66)^2$ & $(208.05)^2$ & $(60.48)^2$ & $(31.14)^2$ \\ 
$0.134$ & $(186.67)^2$ & $(201.47)^2$ & $(177.21)^2$ & $(189.17)^2$ & $(271.89)^2$ & $(97.06)^2$ & $(42.15)^2$ \\ 
$0.183$ & $(263.24)^2$ & $(277.86)^2$ & $(246.38)^2$ & $(260.57)^2$ & $(343.33)^2$ & $(165.26)^2$ & $(59.61)^2$ \\ 
$0.242$ & $(353.31)^2$ & $(372.23)^2$ & $(317.65)^2$ & $(333.16)^2$ & $(428.38)^2$ & $(303.60)^2$ & $(87.07)^2$ \\ 
$0.320$ & $(468.72)^2$ & $(489.32)^2$ & $(450.48)^2$ & $(466.60)^2$ & $(538.21)^2$ & $(442.66)^2$ & $(127.87)^2$ \\ 
$0.430$ & $(640.79)^2$ & $(672.55)^2$ & $(598.53)^2$ & $(618.12)^2$ & $(706.98)^2$ & $(675.26)^2$ & $(185.62)^2$ \\ 
\bottomrule
\end{tabular}
\end{table*}

The resulting power spectra are shown in Fig.~\ref{fig:PS3D_data_model_biascorr} and the values reported in Table~\ref{table:delta_21_ul}. The comparison between $\Delta_{21,\text{d}}^2$ (orange points) and $\Delta_{21,\text{m}}^2$ (grey points) leads to the same conclusions as in Section~\ref{sec:validation:data-model}, where no additional GPR component appeared necessary, except at $k = 0.242\,h\,\text{cMpc}^{-1}$, where $\Delta_{21,\text{m}}^2 < \Delta_{21,\text{d}}^2$. No further investigation was required as the measurements are consistent within the $2\sigma$ uncertainties. Given the potential signal loss indicated by the injection test, we corrected the noise-subtracted power spectra of the residual data by dividing $\Delta_{21,\text{d}}^2$ by the bias factor $b_k$ at each $k$-bin. The bias-corrected power spectra (green points) are higher than $\Delta_{21,\text{d}}^2$ at $k < 0.3\,h\,\text{cMpc}^{-1}$, but they become comparable at smaller scales, where the injection tests mostly returned $z\text{-score} > -2$. For reporting upper limits, we consider the bias-corrected power spectra from the Stokes~I data residuals as our final results, following a conservative approach, so that $\Delta_{21}^2 = \Delta_{21,\text{d}}^2 / b_k$.

The deepest (bias-corrected) upper limit from the 3C\,196 field at $z=9.16$ is $\Delta_{21}^2 < (146.61\,\text{mK})^2$ at $k=0.078\,h\,\text{cMpc}^{-1}$, which is still more than 40 times higher than the achievable upper limit if the residuals were fully consistent with the thermal noise\footnote{Subtracting the noise bias from $\Delta_{\text{I,res}}$, if this is consistent with the thermal noise, the $2\sigma$ upper limits would correspond to the $2\sigma$ uncertainties of the noise.}. This is largely due to the bias correction and the inclusion of the excess variance in the upper limits. At low $k$, the residual power spectrum is dominated by the excess and subtracting the noise bias, which is an order of magnitude lower, does not significantly reduce the upper limits. Conversely, at higher $k$, where the excess power is lower, the residuals are noise-dominated. Here, the difference from the deepest achievable upper limits decreases to a factor of 14 at $k = 0.43\,h\,\text{cMpc}^{-1}$, compared to a factor of 100 at the same $k$ in \citet{mertens_etal:2020}. 


\section{Discussion and conclusions}\label{sec:conclusions}

For the first time, we have used LOFAR observations of the 3C\,196 field to set upper limits on the 21-cm signal from the EoR. Previous LOFAR 21-cm signal limits were all set using the NCP target field. The 3C\,196 field, being closer to the zenith than the NCP, can be observed for only half of the time spent on the NCP but reaches, in principle, the same power spectrum sensitivity. In this work, only a 6-hour night of 3C\,196 data has been processed. The main conclusions of our analysis are the following:

\paragraph*{New 21-cm signal processing pipeline:} The processing is largely
similar between the 3C\,196 and NCP fields, but here, instead of \textsc{sagecal}, we used \dppp, which is the standard tool for processing LOFAR data. The most notable difference between the two lies in the spectral smoothing of the DD-gain solutions. While \textsc{sagecal} employs a consensus optimisation algorithm to spectrally constrain the solutions using a third-order Bernstein polynomial, the \textsc{ddecal} tool of \dppp\ applies a Gaussian smoothing kernel to regularise between iterations. \citet{gan_etal:2023} demonstrated that using a kernel size of 4\,MHz yields power spectrum results comparable to those obtained with \textsc{sagecal} on the same dataset; therefore, we adopted the same setting. 

\paragraph*{Sky modelling and source clustering:} While the NCP model contains more sources than the 3C\,196 sky model, its creation involved significantly more manual work \citep[e.g.][]{bernardi_etal:2010,yatawatta:2013} compared to the automatic procedure used here, which is essentially based on a multi-scale and multi-frequency deconvolution done with \textsc{wsclean} on previously DI calibrated data. Moreover, the NCP model was based on apparent flux densities, whereas in this work, we consistently used intrinsic flux densities. This facilitates the use of our model when the field is observed at different local sidereal times. Another notable difference in this work includes the use of fewer directions in the direction-dependent calibration step (48 for 3C\,196 compared to just over 100 for NCP). 

\paragraph*{Difference in excess power and systematics to NCP field:} Having a second field to compare with opens new insights into the cause of the excess power. One of the most striking results from this work is the different behaviour and level of the excess variance in the 3C\,196 field compared to the NCP field. The excess frequency coherence and baseline behaviour are different from the excess observed in the NCP observations. Compared to the NCP field, the 3C\,196 field has notably different characteristics:
\begin{itemize}
    \item The presence of a very bright FR-II source (i.e.\ 3C\,196) of 83\,Jy at 150\,MHz in the centre of the field, compared to the NCP brightest source of ${\approx}5$\,Jy.
    
    \item It is observed by LOFAR near zenith, leading to higher sensitivity but larger beam variations due to tracking, while the NCP field is observed at a constant zenith angle of $37^\circ$.

    \item Tracking the field may reduce the impact of stationary RFI by time-smearing static interference, while this does not occur in NCP observations, where stationary RFI adds up coherently.
\end{itemize}
A more subtle factor may be the variation of the apparent brightness of sources in the primary beam sidelobes, particularly the four dominant northern A-team sources (i.e.\ Cas\,A, Cyg\,A, Tau\,A and Vir\,A). The combination of all these differences makes it hard to attribute the observed differences to a single cause as of yet, but the excess power in the 3C\,196 results shows a clear wedge-like structure in the cylindrical power spectrum, and an increase in power towards smaller $k$-values. This suggests that its origin is a residual foreground modulated by the time-varying primary beam, rather than, for example, interference.

\paragraph*{Bias factor from signal injection test:} We used GPR for residual foreground removal, which was not used by \citet{patil_etal:2017} but was implemented in more recent NCP analyses \citep{mertens_etal:2020,mertens_etal:2025}. The signal injection test to validate our ML-GPR model returned a median bias factor $b_k$ smaller than one for all the $k$-bins, with $b_k\approx 0.5$ at $k = 0.078\,h\,\text{cMpc}^{-1}$. Such a bias is not measured in the NCP field and may result from a less complete sky model, which covers just the main lobe of the primary beam. In this case, such a bias could be removed by constructing a deeper and more spatially extended sky model.

\paragraph*{New upper limits:} We set a best (bias-corrected) upper limit of $\Delta_{21}^2 < (146.61\,\text{mK})^2$ at $k = 0.078\,h\,\text{cMpc}^{-1}$ and $z = 9.16$ from the 3C\,196 field. This is approximately seven times higher than the current deepest upper limit from the NCP field at the same $k$-bin and redshift \citep{mertens_etal:2025}. However, this difference is expected, given that our result is based on a single 6-hour night, whereas the NCP upper limits are derived from 141 hours of data. A more appropriate comparison is with \citet{patil_etal:2017}, which analysed a single 13-hour night of NCP observations. The longer integration time compensates for the higher noise variance at the NCP, due to its greater distance from the zenith, and thus we expect similar upper limits. Although the $k$-bins do not match exactly, \citet{patil_etal:2017} report an upper limit of $\Delta_{21}^2 < (296.1\,\text{mK})^2$ at $k = 0.1\,h\,\text{cMpc}^{-1}$, whereas our limit at the same $k$ is more than a factor of two lower, with $\Delta_{21}^2 < (208.05\,\text{mK})^2$. This highlights the potential of the 3C\,196 field to provide deeper upper limits than the NCP field, despite our conservative approach of correcting for the GPR bias. Even stronger constraints could be achieved in the future if this bias is reduced or eliminated.

\vspace{\baselineskip} 

In conclusion, our results suggest that the 3C\,196 field has lower systematics and could achieve deeper upper limits than the NCP field when a larger number of nights is processed, potentially providing competitive constraints on the 21-cm signal. One of the advantages of having two independent deep fields is that their power spectra can be combined incoherently, reducing the upper limits by a factor of $\sqrt{2}$. Further comparisons between the two fields will help identify the origin of the excess power and refine observational strategies for future experiments.

\section*{Acknowledgements}

EC (Groningen), ARO and LVEK would like to acknowledge support from the Centre for Data Science and Systems Complexity (DSSC), Faculty of Science and Engineering at the University of Groningen. EC acknowledges support from the Ministry of Universities and Research (MUR) through the PRIN project `Optimal inference from radio images of the epoch of reionization'. LVEK and SM acknowledge the financial support from the European Research Council (ERC) under the European Union's Horizon 2020 research and innovation programme (Grant agreement No.\ 884760, `CoDEX'). FGM acknowledges support from a PSL Fellowship. EC (Nottingham) acknowledges the support of a Royal Society Dorothy Hodgkin Fellowship and a Royal Society Enhancement Award. RG acknowledges support from SERB, DST Ramanujan Fellowship no.~RJF/2022/000141. SKG is supported by NWO grant number OCENW.M.22.307. GM acknowledges support from Swedish Research Council grant 2020-04691. LOFAR, the Low Frequency Array designed and constructed by ASTRON, has facilities in several countries, that are owned by various parties (each with their own funding sources), and that are collectively operated by the International LOFAR Telescope (ILT) foundation under a joint scientific policy. 

Apart from already mentioned software, in this work, we made use of the \textsc{kvis} \citep{gooch:1996} and \textsc{ds9} \citep{joye_mandel:2003} FITS file image viewers, and the \textsc{astropy} \citep{astropy_coll:2022}, \textsc{matplotlib} \citep{hunter:2007}, \textsc{numpy} \citep{harris_etal:2020}, \textsc{pandas} \citep{mckinney:2010}, \textsc{scipy} \citep{virtanen_etal:2020} \textsc{python} packages. 

\section*{Data Availability}

The data underlying this article will be shared on reasonable request to the corresponding author.



\bibliographystyle{mnras}
\bibliography{3c196} 



\appendix

\onecolumn

\section{Direction-dependent sky model}\label{app:dd-skymodel}

In Table~\ref{table:cluster_info}, we provide information on each of the 48 clusters that constitute the 3C\,196 `DD sky model' created in Section~\ref{sec:mod:sky-model-extraction}.

\begin{table*}
\centering
\caption{Details of the 3C\,196 field `DD sky model' used in the EoR processing pipeline. For each cluster, we provide the number of clean components, central coordinates, distance from the field centre, total flux density of the components at 150\,MHz, and the most notable dominating sources. Cluster 48 contains only Cas\,A.}
\label{table:cluster_info}
\begin{tabular}{lcccccl}
\toprule
Cluster & Components & Right Ascension & Declination & Distance & Flux & Bright sources \\
{} & {} & (J2000) & (J2000) & (deg) & (Jy) & {}\\ \midrule
1  & 1864 & $08^\text{h}13^\text{m}35\rlap{.}^\text{s}40$ & $+48^\circ12'52\rlap{.}''15$ & 0.00 & 84.84 & 3C\,196 \\
2  & 205 & $07^\text{h}59^\text{m}28\rlap{.}^\text{s}42$ & $+50^\circ37'59\rlap{.}''52$ & 3.33 & 17.68 & J0801 \\
3  & 105 & $08^\text{h}20^\text{m}11\rlap{.}^\text{s}05$ & $+46^\circ48'59\rlap{.}''74$ & 1.79 & 14.76 & 3C\,197.1 \\
4  & 142 & $08^\text{h}14^\text{m}46\rlap{.}^\text{s}46$ & $+46^\circ32'10\rlap{.}''76$ & 1.69 & 10.76 & 4C+46.17 \\
5  & 148 & $08^\text{h}04^\text{m}45\rlap{.}^\text{s}87$ & $+48^\circ12'42\rlap{.}''03$ & 1.47 & 10.24 & 4C+48.21, J0805 \\
6  & 230 & $08^\text{h}04^\text{m}41\rlap{.}^\text{s}64$ & $+47^\circ14'36\rlap{.}''87$ & 1.78 & 10.17 & 4C+47.27 \\
7  & 232 & $08^\text{h}32^\text{m}01\rlap{.}^\text{s}61$ & $+46^\circ38'33\rlap{.}''93$ & 3.49 & 9.67 & 4C+45.16 \\
8  & 251 & $08^\text{h}21^\text{m}34\rlap{.}^\text{s}60$ & $+51^\circ18'22\rlap{.}''39$ & 3.35 & 9.21 & \\
9  & 319 & $08^\text{h}00^\text{m}13\rlap{.}^\text{s}09$ & $+47^\circ43'30\rlap{.}''54$ & 2.29 & 8.83 & \\
10 & 297 & $08^\text{h}16^\text{m}45\rlap{.}^\text{s}97$ & $+44^\circ52'18\rlap{.}''25$ & 3.39 & 8.03 & \\
11 & 272 & $08^\text{h}02^\text{m}28\rlap{.}^\text{s}17$ & $+46^\circ16'22\rlap{.}''23$ & 2.71 & 7.85 & 4C+45.14 \\
12 & 307 & $08^\text{h}24^\text{m}39\rlap{.}^\text{s}05$ & $+45^\circ24'46\rlap{.}''95$ & 3.38 & 7.61 & \\
13 & 180 & $08^\text{h}05^\text{m}46\rlap{.}^\text{s}53$ & $+49^\circ45'57\rlap{.}''56$ & 2.01 & 7.41 & \\
14 & 300 & $08^\text{h}09^\text{m}12\rlap{.}^\text{s}23$ & $+50^\circ32'58\rlap{.}''73$ & 2.44 & 7.07 & \\
15 & 261 & $08^\text{h}15^\text{m}23\rlap{.}^\text{s}47$ & $+50^\circ22'49\rlap{.}''12$ & 2.19 & 6.99 & \\
16 & 287 & $08^\text{h}10^\text{m}15\rlap{.}^\text{s}24$ & $+45^\circ22'20\rlap{.}''55$ & 2.90 & 6.62 & \\
17 & 168 & $08^\text{h}10^\text{m}11\rlap{.}^\text{s}40$ & $+49^\circ24'27\rlap{.}''33$ & 1.32 & 6.71 & 4C+49.17 \\
18 & 334 & $08^\text{h}14^\text{m}40\rlap{.}^\text{s}70$ & $+51^\circ26'14\rlap{.}''58$ & 3.23 & 6.42 & \\
19 & 260 & $08^\text{h}32^\text{m}51\rlap{.}^\text{s}59$ & $+49^\circ10'00\rlap{.}''74$ & 3.32 & 5.86 & \\
20 & 282 & $08^\text{h}19^\text{m}44\rlap{.}^\text{s}00$ & $+50^\circ00'01\rlap{.}''06$ & 2.05 & 5.79 & \\
21 & 206 & $08^\text{h}19^\text{m}37\rlap{.}^\text{s}25$ & $+45^\circ58'18\rlap{.}''25$ & 2.47 & 5.22 & \\
22 & 289 & $08^\text{h}30^\text{m}41\rlap{.}^\text{s}58$ & $+50^\circ18'17\rlap{.}''88$ & 3.49 & 4.95 & \\
23 & 255 & $08^\text{h}25^\text{m}54\rlap{.}^\text{s}36$ & $+46^\circ32'59\rlap{.}''68$ & 2.67 & 4.74 & \\
24 & 198 & $08^\text{h}27^\text{m}31\rlap{.}^\text{s}48$ & $+48^\circ33'38\rlap{.}''99$ & 2.34 & 4.65 & \\
25 & 183 & $08^\text{h}25^\text{m}53\rlap{.}^\text{s}42$ & $+47^\circ35'37\rlap{.}''47$ & 2.15 & 4.75 & \\
26 & 237 & $08^\text{h}16^\text{m}09\rlap{.}^\text{s}52$ & $+49^\circ21'40\rlap{.}''51$ & 1.22 & 4.67 & \\
27 & 192 & $08^\text{h}02^\text{m}32\rlap{.}^\text{s}20$ & $+48^\circ49'26\rlap{.}''63$ & 1.93 & 4.55 & \\
28 & 296 & $08^\text{h}07^\text{m}45\rlap{.}^\text{s}51$ & $+51^\circ26'56\rlap{.}''72$ & 3.37 & 4.44 & \\
29 & 377 & $08^\text{h}04^\text{m}24\rlap{.}^\text{s}96$ & $+45^\circ10'35\rlap{.}''09$ & 3.42 & 4.11 & \\
30 & 207 & $08^\text{h}21^\text{m}35\rlap{.}^\text{s}04$ & $+49^\circ11'41\rlap{.}''45$ & 1.64 & 4.17 & \\
31 & 125 & $08^\text{h}18^\text{m}55\rlap{.}^\text{s}24$ & $+47^\circ41'38\rlap{.}''58$ & 1.03 & 3.98 & \\
32 & 308 & $07^\text{h}53^\text{m}23\rlap{.}^\text{s}72$ & $+47^\circ32'37\rlap{.}''77$ & 3.45 & 3.96 & \\
33 & 142 & $08^\text{h}14^\text{m}05\rlap{.}^\text{s}45$ & $+47^\circ33'27\rlap{.}''66$ & 0.66 & 3.89 & \\
34 & 288 & $07^\text{h}53^\text{m}30\rlap{.}^\text{s}27$ & $+49^\circ13'14\rlap{.}''77$ & 3.46 & 3.86 & \\
35 & 313 & $08^\text{h}33^\text{m}49\rlap{.}^\text{s}98$ & $+48^\circ03'04\rlap{.}''84$ & 3.38 & 3.77 & \\
36 & 302 & $07^\text{h}56^\text{m}33\rlap{.}^\text{s}12$ & $+48^\circ30'42\rlap{.}''17$ & 2.85 & 3.84 & \\
37 & 120 & $08^\text{h}10^\text{m}42\rlap{.}^\text{s}13$ & $+48^\circ38'39\rlap{.}''80$ & 0.64 & 3.88 & \\
38 & 246 & $08^\text{h}09^\text{m}27\rlap{.}^\text{s}78$ & $+46^\circ32'00\rlap{.}''40$ & 1.82 & 3.23 & \\
39 & 213 & $08^\text{h}00^\text{m}02\rlap{.}^\text{s}18$ & $+49^\circ34'00\rlap{.}''91$ & 2.61 & 3.21 & \\
40 & 186 & $08^\text{h}22^\text{m}29\rlap{.}^\text{s}39$ & $+48^\circ30'05\rlap{.}''16$ & 1.51 & 3.02 & \\
41 & 124 & $08^\text{h}08^\text{m}50\rlap{.}^\text{s}00$ & $+48^\circ09'15\rlap{.}''96$ & 0.80 & 2.70 & \\
42 & 302 & $07^\text{h}57^\text{m}35\rlap{.}^\text{s}33$ & $+46^\circ17'19\rlap{.}''42$ & 3.33 & 2.67 & \\
43 & 248 & $08^\text{h}24^\text{m}32\rlap{.}^\text{s}11$ & $+50^\circ29'43\rlap{.}''99$ & 2.89 & 2.09 & \\
44 & 110 & $08^\text{h}15^\text{m}43\rlap{.}^\text{s}33$ & $+48^\circ34'08\rlap{.}''63$ & 0.50 & 2.03 & \\
45 & 184 & $08^\text{h}09^\text{m}29\rlap{.}^\text{s}40$ & $+47^\circ28'17\rlap{.}''88$ & 1.01 & 1.95 & \\
46 & 202 & $08^\text{h}26^\text{m}12\rlap{.}^\text{s}20$ & $+49^\circ29'50\rlap{.}''70$ & 2.44 & 1.88 & \\
47 & 110 & $08^\text{h}19^\text{m}03\rlap{.}^\text{s}97$ & $+48^\circ20'07\rlap{.}''44$ & 0.93 & 1.35 & \\
48 & 10 & $23^\text{h}23^\text{m}24\rlap{.}^\text{s}23$ & $+58^\circ48'49\rlap{.}''79$ & 66.14 & 11\,734.28 & Cas\,A \\ \bottomrule
\end{tabular}
\end{table*}

\FloatBarrier
\clearpage


\section{Calibration solutions}

\begin{figure*}
    \centering
    \includegraphics[width=0.7\textwidth]{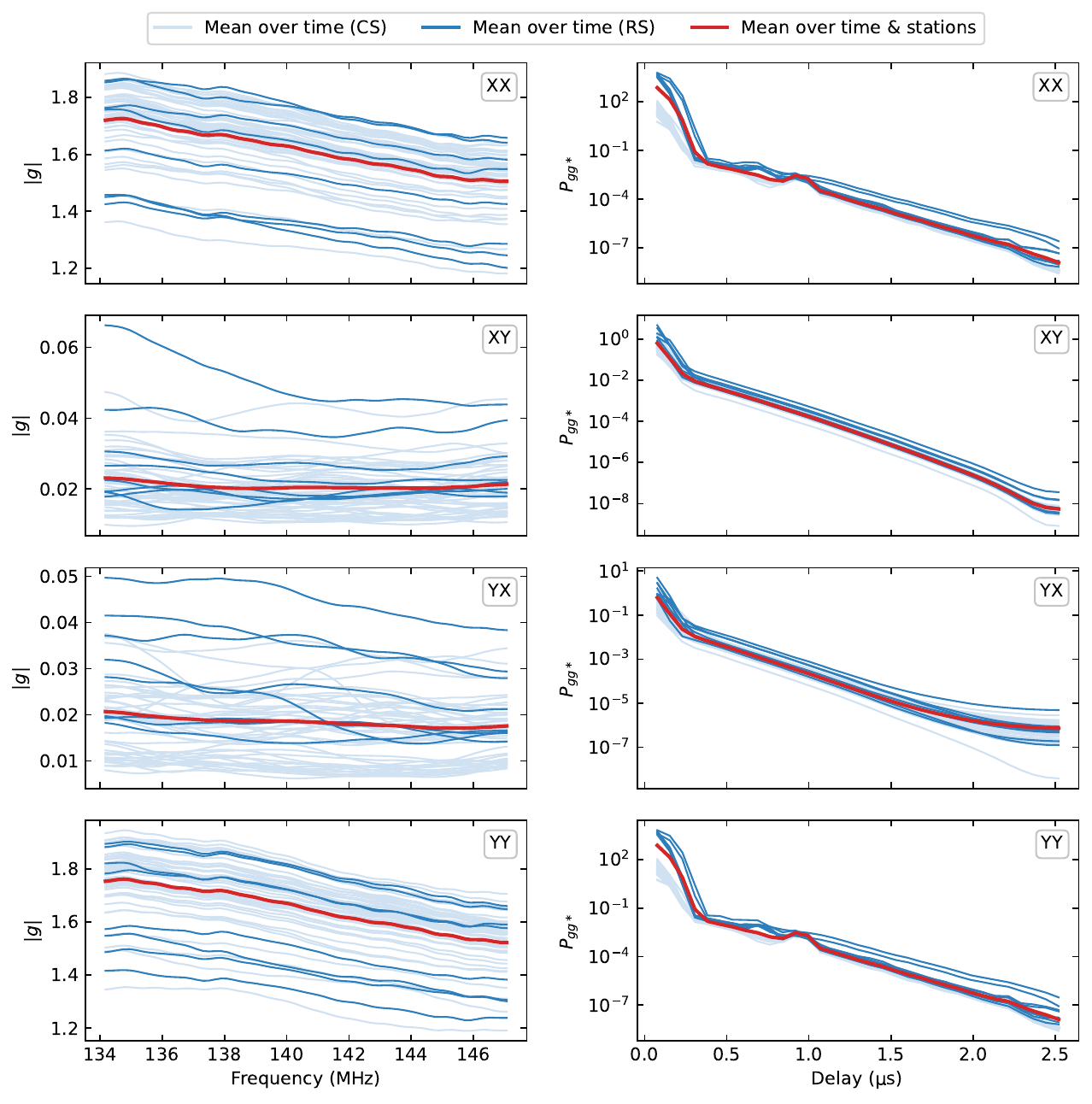}
    \caption{Calibration solutions obtained from the DI spectral smooth calibration of the 3C\,196 field. The amplitude of the station gain $g$ as a function of frequency is shown on the left, and the power spectrum of $gg^*$ as a function of delay is shown on the right. Each row corresponds to one element of the $2 \times 2$ Jones matrix. Time-averaged solutions are shown for each core station (light blue) and remote station (dark blue), with the mean over all stations shown in red.}
    \label{fig:cal_smooth}
\end{figure*}

\begin{figure*}
    \centering
    \includegraphics[width=0.7\textwidth]{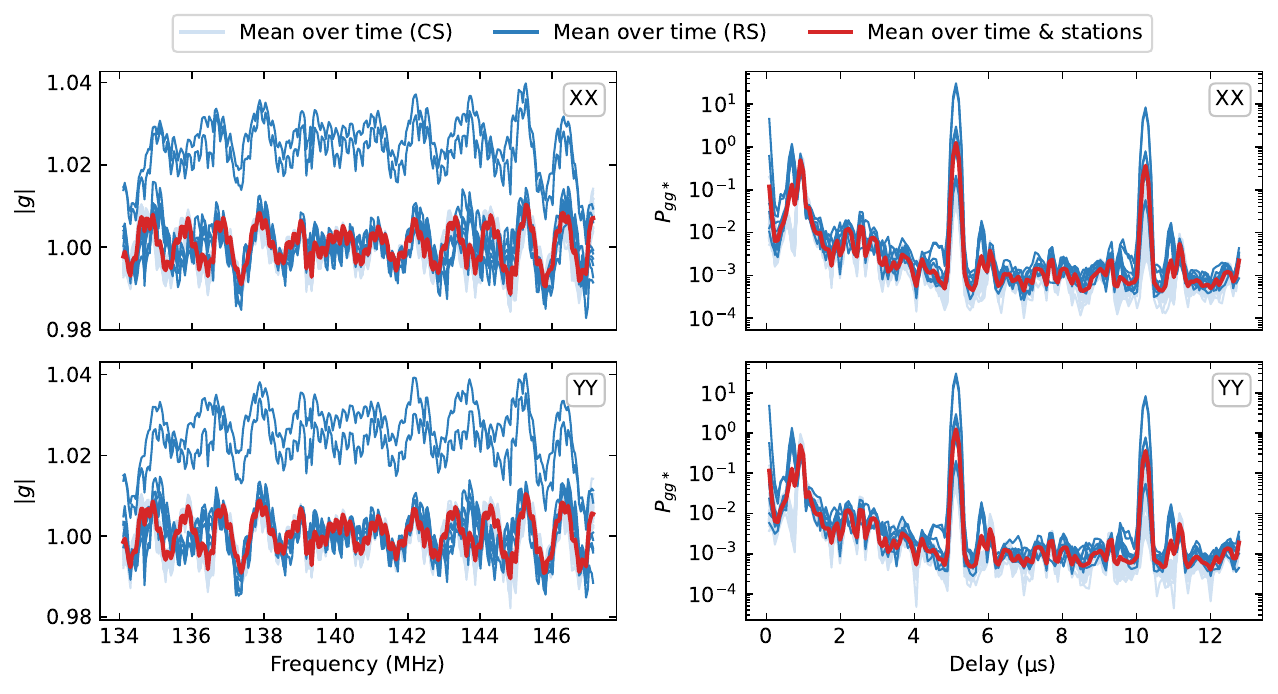}
    \caption{Calibration solutions obtained from the DI bandpass calibration of the 3C\,196 field. The amplitude of the station gain $g$ as a function of frequency is shown on the left, and the power spectrum of $gg^*$ as a function of delay is shown on the right. The XX and YY elements of the Jones matrix are shown in the top and bottom rows, respectively. Time-averaged solutions are shown for each core station (light blue) and remote station (dark blue), with the mean over all stations shown in red.}
    \label{fig:cal_bp}
\end{figure*}

\begin{figure*}
    \centering
    \includegraphics[width=0.7\textwidth]{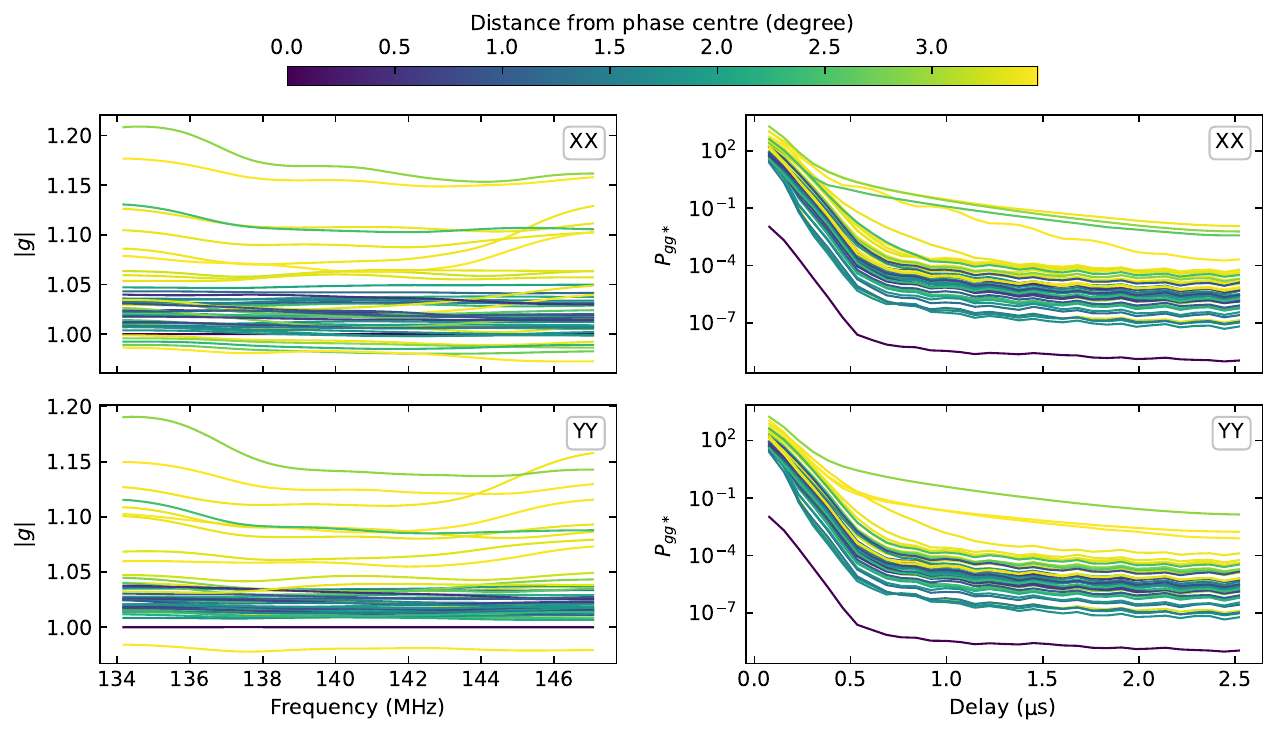}
    \caption{Calibration solutions obtained from the DD calibration of the 3C\,196 field. The amplitude of the gain $g$ as a function of frequency is shown on the left, and the power spectrum of $gg^*$ as a function of delay is shown on the right. The XX and YY elements of the Jones matrix are shown in the top and bottom rows, respectively. Each line represents the time and station-averaged gains for each of the 47 main field clusters. The cluster distance from the field centre is colour-coded, ranging from dark blue for 3C\,196 to yellow for the farthest cluster ($3.49^\circ$).}
    \label{fig:cal_dd}
\end{figure*}

\begin{figure*}
    \centering
    \includegraphics[width=0.7\textwidth]{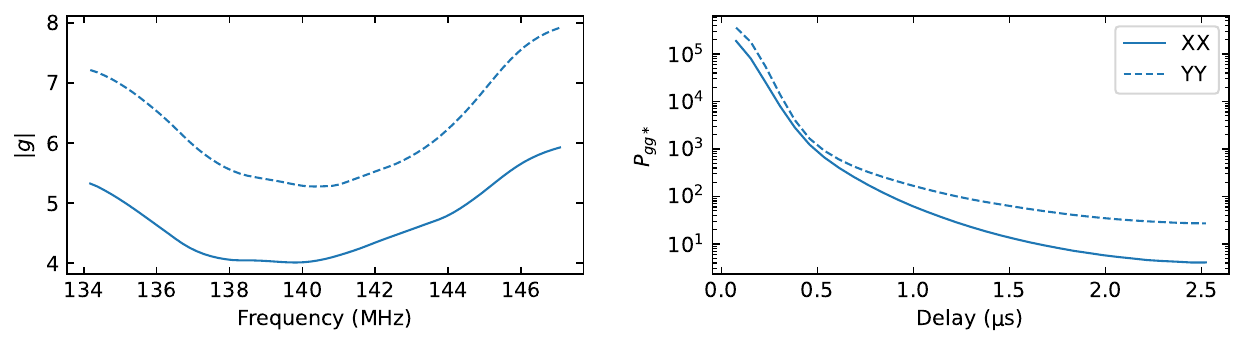}
    \caption{Calibration solutions obtained from the DD calibration of the 3C\,196 field for the Cas\,A direction. The amplitude of the gain $g$ as a function of frequency is shown on the left, and the power spectrum of $gg^*$ as a function of delay is shown on the right. The XX and YY elements of the Jones matrix are shown with solid and dashed lines, respectively. The solutions are averaged over time and stations.}
    \label{fig:cal_dd_casa}
\end{figure*}

Here, we provide a summary of the calibration solutions obtained from running the EoR pipeline on our 3C\,196 field dataset (see Section~\ref{sec:3c196-processing}). In each figure, we plot the amplitude of the station gains $g$ as a function of frequency (left) and the power spectrum of $gg^*$ as a function of time delay (right). Figures~\ref{fig:cal_smooth} and \ref{fig:cal_bp} show the time-averaged solutions from the DI spectral smooth and bandpass calibrations, respectively. The DD calibration solutions, averaged over time and stations, are shown in Fig.~\ref{fig:cal_dd} for each of the main field clusters, with different colours indicating the cluster distance from the centre (see Table~\ref{table:cluster_info}). The DD solutions for the Cas\,A direction (cluster 48) are presented separately in Fig.~\ref{fig:cal_dd_casa}.

\section{Linearly polarized intensity map}\label{app:polQU}

\begin{figure*}
    \centering
    \includegraphics[width=0.95\textwidth]{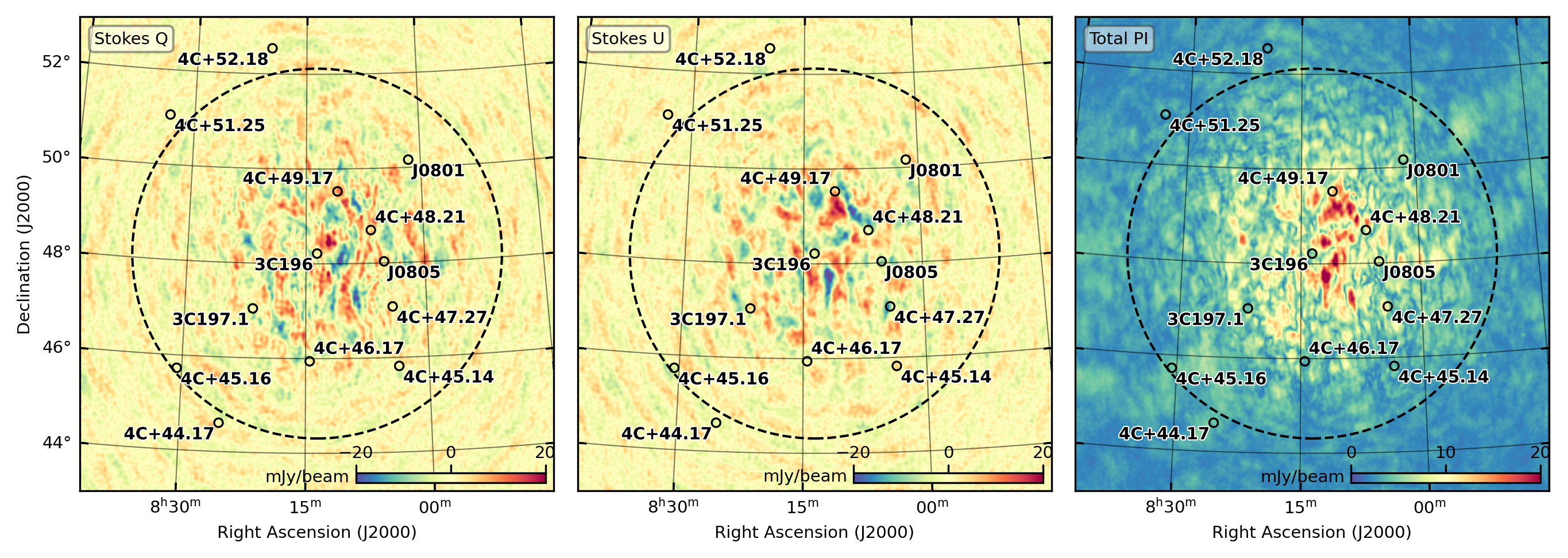}
    \caption{Stokes~Q (left) and U (middle) images at 140.4\,MHz of the 3C\,196 field. The total PI map at a Faraday depth of $0\,\text{rad/m}^2$ is shown in the right panel. All the images have a pixel scale of 0.5\,arcmin and a field of view of $10^\circ\times 10^\circ$. The $u\varv$-plane was gridded with a natural weighting scheme and only baselines between 50 and $500\lambda$ were used. The black dashed circle indicates the $3.9^\circ$ radius of the sky model. The brightest sources are highlighted with black and red circles.}
    \label{fig:3C196_afterDD_StokesQ_U}
\end{figure*}

The left and middle panels of Fig.~\ref{fig:3C196_afterDD_StokesQ_U} show the Stokes~Q and U images at 140.4\,MHz of the 3C\,196 field after the DD subtraction (Section~\ref{sec:proc:dd-cal}). To compare our polarized emission results with \citet{jelic_etal:2015}, we calculate the total polarized intensity (PI) at a Faraday depth of $0\,\text{rad/m}^2$ as
\begin{equation} 
\text{PI}= \frac{1}{N_{\text{SB}}}\sum_{i=1}^{N_\text{SB}} \sqrt{\text{Q}_i^2 + \text{U}_i^2}\, , 
\end{equation}
where $N_\text{SB}$ is the number of individually imaged sub-bands, which in our case is 67 for the redshift bin of interest (corresponding to the 134.2--147.1\,MHz frequency range). Here, $\text{Q}_i$ and $\text{U}_i$ represent the images of each individually imaged sub-band. The obtained PI map only slightly resembles the structures found by \citet{jelic_etal:2015} in Fig.~3. However, a direct comparison is not straightforward due to two key differences: (i) we did not perform any rotation measure (RM) calibration or synthesis, whereas \citet{jelic_etal:2015} corrected for RM using ionospheric data; and (ii) their analysis used a different $u\varv$-cut ($10{-}800\lambda$) compared to ours ($50{-}500\lambda$), leading to differences in the synthesised beam, sensitivity, and scale of emission captured. Nevertheless, we can still conclude that the 3C\,196 field is strongly polarized, and Stokes~Q and U may leak power into Stokes~V, as observed in Fig.~\ref{fig:3C196_afterDD_StokesI_V}.

\section{Tests for excess power kernels}\label{app:excess_tests}

As described in Section~\ref{sec:gpr:model}, most of the effort in constructing the GPR model focused on finding the optimal shape for the excess covariance function. Table~\ref{table:mr_combined} summarises the results from tests with different kernel shapes and hyper-parameter priors. Test 7 gave the highest marginal likelihood and was selected for modelling the residual data. Although Test 3 provided an evidence $\log \mathcal{Z}$ higher than Test 7, all the RBF covariances resulted in converging 21-cm signal parameters. This behaviour, unexpected for a component well below the thermal noise level, suggested potential errors in the model fitting. Therefore, we discarded all the RBF tests, but kept them in the table for completeness.

\begin{table*}
\centering
\caption{Summary of kernel shape tests for the excess component in the ML-GPR model for the 3C\,196 field. Each test explores different covariance functions, parametrizations, and hyper-parameter priors, as described in Section~\ref{sec:gpr:model}. The parameter $\log\mathcal{Z}$ represents the marginal likelihood (evidence), with higher values indicating better model fit.}
\label{table:mr_combined}
\resizebox{1\textwidth}{!}{
\begin{tabular}{lccccccc}
\toprule
Parameter & Test 1 & Test 2 & Test 3 & Test 4 & Test 5 & Test 6 & Test 7 \\
\midrule
Covariance & RBF & RBF & RBF & $\mu=3/2$ & $\mu=5/2$ & $\mu=5/2$ & $\mu=5/2$ \\
Parametrization & -- & Wedge & $\alpha$-coefficient & $\alpha$-coefficient & $\alpha$-coefficient & $\alpha$-coefficient & $\alpha$-coefficient \\
Prior $\sigma_\text{ex}^2 / \sigma_{\text{d}}^2$ & $\log\mathcal{U}(-4, -2)$ & $\log\mathcal{U}(-4, -2)$ & $\log\mathcal{U}(-4, -2)$ & $\log\mathcal{U}(-4, -2)$ & $\log\mathcal{U}(-4, -2)$ & $\log\mathcal{U}(-4, -2)$ & $\log\mathcal{U}(-4, -2)$ \\
Prior $l_\text{ex}$ (MHz) & $\mathcal{U}(0.2, 0.8)$ & -- & -- & -- & -- & -- & --  \\
Prior $\eta_\text{buffer,ex}$ ({\textmu}s) & -- & Fixed(2.6) & -- & -- & -- & -- & -- \\
Prior $\theta_\text{ex}$ (rad) & -- & $\mathcal{U}(0.1, 2)$ & -- & -- & -- & -- & -- \\
Prior $l_{0,\text{ex}}$ (MHz) & -- & -- & $\mathcal{U}(0.2, 0.8)$ & $\mathcal{U}(0.2, 0.8)$ & $\mathcal{U}(0.2, 0.8)$ & $\mathcal{U}(0.2, 0.8)$ & $\mathcal{U}(0.2, 0.8)$ \\
Prior $\alpha_\text{ex}$ & -- & -- & $\mathcal{U}(-5, 60)$ & $\mathcal{U}(-5, 60)$ & $\mathcal{U}(-5, 60)$ & $\mathcal{U}(-5, 60)$ & $\mathcal{U}(-5, 60)$ \\
Prior $\sigma_{\alpha,\text{ex}}^2$ & -- & -- & $\mathcal{U}(-1,1)$ & $\mathcal{U}(-1,1)$ & $\mathcal{U}(-1,1)$ & Fixed(0) & Fixed($-0.25$) \\
$\log\mathcal{Z}$ & $37\,160.6 \pm 0.7$ & $37\,229.5 \pm 0.6$ & $37\,320.5 \pm 0.7$ & $37\,301.3 \pm 0.8$ & $37\,310.7 \pm 0.4$ & $37\,305.2 \pm 0.9$ & $37\,312.8 \pm 0.6$ \\
\bottomrule
\end{tabular}}
\end{table*}


\bsp	
\label{lastpage}
\end{document}